\documentclass[12pt]{article}

\usepackage{graphics,graphicx}
\usepackage{amssymb,epsfig,amsmath,euscript,array}
\usepackage[nosort]{cite}
\usepackage{axodraw4j}
\usepackage{pstricks}         
\usepackage{color}
\usepackage{bbm}

\makeatletter
\@addtoreset{equation}{section}
\makeatother

\newcounter{multieqs}

\newcommand{\be}{\begin{equation}}
\newcommand{\ee}{\end{equation}}

\newcommand{\bm}[1]{\mbox{\boldmath $#1$}}

\newcommand{\kslash}{k \!\!\! / }

\newcommand{\lslash}{l \!\! / }
\newcommand{\Pslash}{P \!\!\!\! / }

\newcommand{\islash}{i \!\!\! / }
\newcommand{\jslash}{j \!\!\! / }
\newcommand{\aslash}{a \!\!\! / }
\newcommand{\bslash}{{b \hspace{-6pt} \slash} }

\newcommand{\onslash}{1 \!\!\! / }
\newcommand{\twslash}{2 \!\!\!/ }
\newcommand{\thslash}{3 \!\!\!/ }
\newcommand{\foslash}{4 \!\!\! / }
\newcommand{\fislash}{5 \!\!\! / }

\newcommand{\mslash}{m \!\!\! / }

\def\bd{\begin{document}}
\def\ed{\end{document}}
\def\nn{\nonumber}
\def\bea{\begin{eqnarray}}
\def\eea{\end{eqnarray}}

\def\ab{(ijab)}
\def\ba{(ijba)}
\def\ijab{{\tr}_{+}(\islash\, \jslash\, \aslash \, \bslash)}
\def\ijba{{\tr}_{+}(\islash\, \jslash\, \bslash \, \aslash)}
\def\ijaP{{\tr}_{+}(\islash\, \jslash\, \aslash \, \Pslash)}
\def\ijPLa{{\tr}_{+}(\islash\, \jslash\, \Pslash_L \, \aslash)}
\def\ijaPL{{\tr}_{+}(\islash\, \jslash\, \aslash \, \Pslash_L)}
\def\ijPLza{{\tr}_{+}(\islash\, \jslash\, \Pslash_{L;z} \, \aslash)}
\def\ijaPLz{{\tr}_{+}(\islash\, \jslash\, \aslash \, \Pslash_{L;z})}
\def\ijPa{{\tr}_{+}(\islash\, \jslash\, \Pslash \, \aslash)}
\def\iaPb{{\tr}_{+}(\islash\, \aslash\, \Pslash \, \bslash)}
\def\ibPa{{\tr}_{+}(\islash\, \bslash\, \Pslash \, \aslash)}
\def\ijPmu{{\tr}_{+}(\islash\, \jslash\, \Pslash \, \mu)}
\def\ibmuP{{\tr}_{+}(\islash\, \bslash\, \mu \, \Pslash)}
\def\ibmua{{\tr}_{+}(\islash\, \bslash\, \mu \, \aslash)}
\def\iamub{{\tr}_{+}(\islash\, \aslash\, \mu \, \bslash)}
\def\jaPb{{\tr}_{+}(\jslash\, \aslash\, \Pslash \, \bslash)}
\def\ijmuP{{\tr}_{+}(\islash\, \jslash\, \mu \, \Pslash)}
\def\ijmum{{\tr}_{+}(\islash\, \jslash\, \mu \, \mslash)}
\def\ijmmu{{\tr}_{+}(\islash\, \jslash\, \mslash \, \mu)}
\def\ijmP{{\tr}_{+}(\islash\, \jslash\, \mslash \, \Pslash)}
\def\iabP{{\tr}_{+}(\islash\, \aslash\, \bslash \, \Pslash)}
\def\ijbP{{\tr}_{+}(\islash\, \jslash\, \bslash \, \Pslash)}
\def\jbPa{{\tr}_{+}(\jslash\, \bslash\, \Pslash \, \aslash)}
\def\ijPb{{\tr}_{+}(\islash\, \jslash\, \Pslash \, \bslash)}
\def\jbmua{{\tr}_{+}(\jslash\, \bslash\, \mu \, \aslash)}

\def\loablt{ {\tr}_{+}(\lslash_1\, \aslash \, \bslash\, \lslash_2)}

\def\ijlolt{{\tr}_{+}(\islash\, \jslash\, \lslash_1 \, \lslash_2)}
\def\ijltlo{{\tr}_{+}(\islash\, \jslash\, \lslash_2 \, \lslash_1)}
\def\ibloa{{\tr}_{+}(\islash\, \bslash\, \lslash_1 \, \aslash)}
\def\jaltb{{\tr}_{+}(\jslash\, \aslash\, \lslash_2 \, \bslash)}
\def\ialtb{{\tr}_{+}(\islash\, \aslash\, \lslash_2 \, \bslash)}
\def\bltloa{{\tr}_{+}(\bslash\, \lslash_2\, \lslash_1 \, \aslash)}
\def\jbloa{{\tr}_{+}(\jslash\, \bslash\, \lslash_1 \, \aslash)}
\def\ibPb{{\tr}_{+}(\islash\, \bslash\, \Pslash \, \bslash)}
\def\ijltb{{\tr}_{+}(\islash\, \jslash\, \lslash_2 \, \bslash)}

\def\ijloa{{\tr}_{+}(\islash\, \jslash\,  \lslash_1 \, \aslash)}
\def\ijblt{{\tr}_{+}(\islash\, \jslash\,  \bslash \, \lslash_2)}

\def\jakb{{\tr}_{+}(\jslash\, \aslash\, \kslash \, \bslash)}
\def\iakb{{\tr}_{+}(\islash\, \aslash\, \kslash \, \bslash)}

\def\tofo{{\tr}_{+}(\onslash\, \thslash\, \twslash \, \foslash)}
\def\foto{{\tr}_{+}(\onslash\, \thslash\, \foslash \, \twslash)}
\def\tofi{{\tr}_{+}(\onslash\, \thslash\, \twslash \, \fislash)}
\def\fito{{\tr}_{+}(\onslash\, \thslash\, \fislash \, \twslash)}

\def\lrangle#1#2{\langle #1\,#2\rangle}

\def\Li{{$\rm Li}_2$}
\def\eps{\epsilon}
\def\epsuv{{\epsilon_{\rm \mbox{\tiny UV}}}}
\let\bm=\bibitem
\let\la=\label

\def\npb#1#2#3{Nucl. Phys. {\bf{B#1}} #3 (#2)}
\def\plb#1#2#3{Phys. Lett. {\bf{#1B}} #3 (#2)}
\def\prl#1#2#3{Phys. Rev. Lett. {\bf{#1}} #3 (#2)}
\def\prd#1#2#3{Phys. Rev. {D \bf{#1}} #3 (#2)}
\def\cmp#1#2#3{Comm. Math. Phys. {\bf{#1}} #3 (#2)}
\def\cqg#1#2#3{Class. Quantum Grav. {\bf{#1}} #3 (#2)}
\def\nppsa#1#2#3{Nucl. Phys. B (Proc. Suppl.) {\bf{#1A}}#3 (#2)}
\def\ap#1#2#3{Ann. of Phys. {\bf{#1}} #3 (#2)}
\def\ijmp#1#2#3{Int. J. Mod. Phys. {\bf{A#1}} #3 (#2)}
\def\rmp#1#2#3{Rev. Mod. Phys. {\bf{#1}} #3 (#2)}
\def\mpla#1#2#3{Mod. Phys. Lett. {\bf A#1} #3 (#2)}
\def\jhep#1#2#3{J. High Energy Phys. {\bf #1} #3 (#2)}
\def\atmp#1#2#3{Adv. Theor. Math. Phys. {\bf #1} #3 (#2)}
\newcommand{\EQ}[1]{\begin{equation} #1 \end{equation}}
\newcommand{\AL}[1]{\begin{subequations}\begin{align} #1 \end{align}\end{subequations}}
\newcommand{\SP}[1]{\begin{equation}\begin{split} #1 \end{split}\end{equation}}
\newcommand{\ALAT}[2]{\begin{subequations}\begin{alignat}{#1} #2 \end{alignat}
                        \end{subequations}}
\def\beqa{\begin{eqnarray}}
\def\eeqa{\end{eqnarray}}
\def\beq{\begin{equation}}
\def\eeq{\end{equation}}
\def\sst{\scriptscriptstyle}
\def\thetabar{\bar\theta}
\def\Tr{{\rm Tr}}
\def\one{\mbox{1 \kern-.59em {\rm l}}}
 \def\Nh{\hat{N}}

\newcommand{\half}{{\textstyle {1 \over 2}}}

\def\a{\alpha}      \def\da{{\dot\alpha}}
\def\b{\beta}       \def\db{{\dot\beta}}
\def\c{\gamma}  \def\G{\Gamma}  \def\cdt{\dot\gamma}
\def\d{\delta}  \def\D{\Delta}  \def\ddt{\dot\delta}
\def\e{\epsilon}        \def\vare{\varepsilon}
\def\f{\phi}    \def\F{\Phi}    \def\vvf{\f}
\def\h{\eta}
\def\k{\kappa}
\def\l{\lambda} \def\L{\Lambda}
\def\m{\mu} \def\n{\nu}
\def\o{\omega}
\def\p{\pi} \def\P{\Pi}
\def\r{\rho}
\def\s{\sigma}  \def\S{\Sigma}
\def\t{\tau}
\def\th{\theta} \def\Th{\Theta} \def\vth{\vartheta}
\def\X{\Xeta}
\def\z{\zeta}
\def\de{\partial}

\def\cA{{\cal A}} \def\cB{{\cal B}} \def\cC{{\cal C}}
\def\cD{{\cal D}} \def\cE{{\cal E}} \def\cF{{\cal F}}
\def\cG{{\cal G}} \def\cH{{\cal H}} \def\cI{{\cal I}}
\def\cJ{{\cal J}} \def\cK{{\cal K}} \def\cL{{\cal L}}
\def\cM{{\cal M}} \def\cN{{\cal N}} \def\cO{{\cal O}}
\def\cP{{\cal P}} \def\cQ{{\cal Q}} \def\cR{{\cal R}}
\def\cS{{\cal S}} \def\cT{{\cal T}} \def\cU{{\cal U}}
\def\cV{{\cal V}} \def\cW{{\cal W}} \def\cX{{\cal X}}
\def\cY{{\cal Y}} \def\cZ{{\cal Z}}

\def\ua{\underline{\alpha}}
\def\ub{\underline{\phantom{\alpha}}\!\!\!\beta}
\def\uc{\underline{\phantom{\alpha}}\!\!\!\gamma}
\def\um{\underline{\mu}}
\def\ud{\underline\delta}
\def\ue{\underline\epsilon}
\def\una{\underline a}\def\unA{\underline A}
\def\unb{\underline b}\def\unB{\underline B}
\def\unc{\underline c}\def\unC{\underline C}
\def\und{\underline d}\def\unD{\underline D}
\def\une{\underline e}\def\unE{\underline E}
\def\unf{\underline{\phantom{e}}\!\!\!\! f}\def\unF{\underline F}
\def\unm{\underline m}\def\unM{\underline M}
\def\unn{\underline n}\def\unN{\underline N}
\def\unp{\underline{\phantom{a}}\!\!\! p}\def\unP{\underline P}
\def\unq{\underline{\phantom{a}}\!\!\! q}
\def\unQ{\underline{\phantom{A}}\!\!\!\! Q}
\def\unH{\underline{H}}

\def\As {{A \hspace{-6.4pt} \slash}\;}
\def\bs {{b \hspace{-6.4pt} \slash}\;}
\def\Ds {{D \hspace{-6.4pt} \slash}\;}
\def\ds {{\del \hspace{-6.4pt} \slash}\;}
\def\ss {{\s \hspace{-6.4pt} \slash}\;}
\def\ks {{ k \hspace{-6.4pt} \slash}\;}
\def\ps {{p \hspace{-6.4pt} \slash}\;}
\def\pas {{{p_1} \hspace{-6.4pt} \slash}\;}
\def\pbs {{{p_2} \hspace{-6.4pt} \slash}\;}
\def\Ps {{P \hspace{-6.4pt} \slash}\;}
\def\Qs {{Q \hspace{-6.4pt} \slash}\;}

\def\Fh{\hat{F}}
\def\Vh{\hat{V}}
\def\Xh{\hat{X}}
\def\ah{\hat{a}}
\def\xh{\hat{x}}
\def\yh{\hat{y}}
\def\ph{\hat{p}}
\def\xih{\hat{\xi}}
\def\psit{\tilde{\psi}}
\def\Psit{\tilde{\Psi}}
\def\tht{\tilde{\th}}
\def\lt{\tilde{\lambda}}
\def\hl{\hat{\lambda}}
\def\hlt{\hat{\tilde{\lambda}}}
\def\llt{\tilde{l}}
\def\At{\tilde{A}}
\def\Qt{\tilde{Q}}
\def\Rt{\tilde{R}}
\def\Nt{\tilde{N}}

\def\at{\tilde{a}}
\def\st{\tilde{s}}
\def\ft{\tilde{f}}
\def\pt{\tilde{p}}
\def\qt{\tilde{q}}
\def\vt{\tilde{v}}
\def\nt{\tilde{n}}

\def\delb{\bar{\partial}}
\def\bz{\bar{z}}
\def\bD{\bar{D}}
\def\bB{\bar{B}}

\def\bk{{\bf k}}
\def\bl{{\bf l}}
\def\bp{{\bf p}}
\def\bq{{\bf q}}
\def\br{{\bf r}}
\def\bx{{\bf x}}
\def\by{{\bf y}}
\def\bR{{\bf R}}
\def\bV{{\bf V}}

\def\d{\delta}\def\D{\Delta}\def\ddt{\dot\delta}
\def\pa{\partial} \def\del{\partial}
\def\xx{\times}
\def\uno{\mbox{1 \kern-.59em {\rm l}}}
\def\trp{^{\top}}
\def\inv{^{-1}}
\def\dag{{^{\dagger}}}
\def\pr{^{\prime}}
\def\lan{\langle}
\def\ran{\rangle}
\def\rar{\rightarrow}
\def\lar{\leftarrow}
\def\lrar{\leftrightarrow}
\newcommand{\0}{\,\!}      
\def\one{1\!\!1\,\,}
\def\im{\imath}
\def\jm{\jmath}
\newcommand{\tr}{\mbox{tr}}
\newcommand{\slsh}[1]{/ \!\!\!\! #1}
\def\vac{|0\rangle}
\def\lvac{\langle 0|}
\def\hlf{\frac{1}{2}}
\def\ove#1{\frac{1}{#1}}
\def\Box{\square}
\def\ZZ{\mathbb{Z}}
\def\CC#1{({\bf #1})}
\def\bcomment#1{}
\def\bfhat#1{{\bf \hat{#1}}}
\def\VEV#1{\left\langle #1\right\rangle}
\newcommand{\ex}[1]{{\rm e}^{#1}} \def\ii{{\rm i}}
\def\rr{{\rm r}} \def\rs{{\rm s}}\def\rv{{\rm v}}
\def\ri{{\rm i}}\def\rj{{\rm j}}
\newcommand{\lrbrk}[1]{\left(#1\right)}
\newcommand{\sfrac}[2]{{\textstyle\frac{#1}{#2}}}

\def\Li{{\rm Li}_2}

\font\mybb=msbm10 at 12pt
\def\bb#1{\hbox{\mybb#1}}

\font\myBB=msbm10 at 18pt
\def\BB#1{\hbox{\myBB#1}}

\begin{document}

\begin{flushright}
QMUL-PH-13-16 \\
ROM2F/2013/17
\end{flushright}

\vspace{8pt}

\begin{center}

{\Large \bf Simplifying instanton corrections to  }
\\
\vspace{0.4cm}
{\Large \bf    $\cN=4$ SYM correlators  }

\vspace{16pt}

{\mbox {\bf  Massimo Bianchi$^{a, b}$, Andreas Brandhuber$^{b}$,  Gabriele Travaglini$^{b}$ and   Congkao Wen$^{b}$}}%
\footnote{
{\tt  massimo.bianchi@roma2.infn.it,\{ \tt \!\!\!a.brandhuber, g.travaglini, c.wen\}@qmul.ac.uk}
}


\begin{quote}
{\small \em
\begin{itemize}
\item[\ \ \ \ \ \ $^a$]
Dipartimento di Fisica, 
Universit\`a di Roma ``Tor Vergata"  \\
\& I.N.F.N. Sezione di Roma ``Tor Vergata" \\
Via della Ricerca Scientifica, 00133 Roma, Italy
\item[\ \ \ \ \ \ $^b$]
\begin{flushleft}
Centre for Research in String Theory\\
School of Physics and Astronomy\\
Queen Mary University of London\\
Mile End Road, London E1 4NS, United Kingdom
\end{flushleft}
\end{itemize}
}
\end{quote}


\vspace{60pt} {\bf Abstract}
\end{center}

\noindent
We compute instanton corrections to non-minimal correlation functions of chiral primary operators in the $\cN=4$ super Yang-Mills super-current multiplet.
Using a representation in terms of Mellin integrals, we find that these corrections can be written as conformal integrals in AdS$_5$ times certain  ``kinematic" prefactors that are independent of the instanton moduli. We then consider the consecutive, pairwise light-like limit $x_{i, i+1}^2 \to 0$ of such correlators, and  prove that the ratio between the instanton contribution and the corresponding tree-level expression vanishes in this limit. 
We also speculate on the extension to the non-perturbative level of the proposed duality with polygonal Wilson loops and briefly discuss its potential implications for scattering amplitudes.

\setcounter{page}{0}
\thispagestyle{empty}
\newpage


\setcounter{tocdepth}{4}
\hrule height 0.75pt
\tableofcontents
\vspace{0.8cm}
\hrule height 0.75pt
\vspace{1cm}

\setcounter{tocdepth}{2}


\setcounter{footnote}{0}

\section{Introduction and motivations}
\label{intromotiv}

The holographic correspondence between $\mathcal{N}=4$ super Yang-Mills (SYM) in four dimensions and Type IIB superstrings on AdS$_5\times S^5$ has passed many tests and brought unprecedented progress in our understanding of conformal field theories (CFT's) in $D=4$, and of the dynamics of strings and gravity. In the planar limit, both sides of the correspondence seem to be integrable, thus allowing for a systematic analysis of the spectrum of anomalous dimensions, dual to masses of string states in AdS, that together with the operator product expansion (OPE) coefficients represent the basic observables of any CFT. 

Other observables have received increasing attention in the literature thanks to their rather surprising interconnections. In particular, a certain class of scattering amplitudes, known as Maximally Helicity Violating (MHV), has been shown to be related to the expectation value of polygonal, light-like Wilson loops \cite{Alday:2007hr, Drummond:2007aua, Brandhuber:2007yx}. 
More recently an  intriguing relation between correlation functions in a particular light-like limit and light-like Wilson loops (in the adjoint representation) has been proposed \cite{Alday:2010zy, Eden:2010zz}, which then leads to a triality between scattering amplitudes, Wilson loops and correlation functions. This triality has been further extended to full superamplitudes in $\mathcal{N}=4$ SYM by considering super Wilson loops and super correlation functions \cite{Mason:2010yk, CaronHuot:2010ek, Eden:2011yp, Eden:2011ku}.

It is well known that  correlation functions in $\mathcal{N}=4$ SYM  receive non-perturbative, instanton corrections \cite{Bianchi:1998nk,Dorey:1999pd}, which are the counterpart of D-instanton corrections to higher-derivative terms in the  type IIB superstring effective action 
\cite{Green:1997tv,Banks:1998nr}.  
With this in mind, it is natural to wonder whether light-like Wilson loops and scattering amplitudes may receive  instanton corrections as well. 
In this paper we will show that at  leading order in $g^2_{\rm YM}$, instanton contributions to correlation functions vanish in the appropriate  light-like limit, after dividing by the corresponding tree-level correlator. Here the relevant limit in question is the one which in perturbation theory leads to the duality with light-like polygonal Wilson loops.  
In \cite{Alday:2010zy} a perturbative proof of this duality  was presented that does not rely on taking the large-$N$ limit.  
Our result and the proof of \cite{Alday:2010zy}  make it plausible that the correlation function/Wilson loop duality holds non-perturbatively, and imply that instanton corrections to light-like Wilson loops are  absent at  leading order in $g^2_{\rm YM}$.%
\footnote{As for the correlation function/amplitude duality, one has to be more careful as it requires the 't Hooft large-$N$ limit. A  discussion of the large-$N$ limit we are considering in this paper will be presented later in this Introduction.} 

This result is perhaps not unexpected since it is known that Konishi-like operators do not receive instanton corrections to their anomalous dimension, at least at  leading order in $g^2_{\rm YM}$, and the same should hold for higher-spin operators associated to string states and their Kaluza-Klein (KK) excitations,  which decouple from supergravity at strong coupling (large AdS radius). 
We also note that an indication of instanton contributions to scattering amplitudes  would be the presence of instanton corrections to the cusp anomalous dimension -- a quantity that controls the  ultraviolet divergences of cusped Wilson loops and the infrared divergences of amplitudes \cite{Polyakov:1980ca,Brandt:1981kf,Korchemsky:1985xj}. 
In the planar limit and at strong coupling, it is known that  there are non-perturbative corrections  to the cusp anomalous dimension \cite{Basso:2007wd},  
which however should admit a completely different explanation, as evident from their scaling with 
$\lambda^{1 / 4} e^{-\sqrt{\lambda}/2}$, with $\lambda:=g^2_{\rm YM} N$,   rather than $e^{-8\pi^2 / g_{\rm YM}^2}$. For completeness, we mention that the class of operators that are known to receive instanton corrections to their anomalous dimension consists of the non-protected multi-trace operators that appear in the OPE of CPO's and their super-descendents.

Of course, an alternative and more  direct way to prove the absence/presence of non-perturbative corrections to amplitudes would consist in applying the LSZ reduction to   correlation functions of fundamental fields evaluated in an instanton background. Focusing for instance on MHV amplitudes, it is easy to see that, to lowest order, the instanton contribution to the  corresponding correlation functions  vanishes since the exact fermionic zero-modes around the instanton cannot be absorbed.  
We will sketch the strategy of this alternative approach in Section \ref{direct}, postponing a more thorough analysis, in particular of potential subtle infrared  singularities, to \cite{LSZ}. It is however interesting to point out that such subtleties are absent  in the case of amplitudes in (non conformal) theories with lower supersymmetry. A case in point is  that of $\mathcal{N}=2$ SYM \cite{SW}, where instanton corrections to the pre-potential in the Coulomb branch can be related to scattering amplitudes of gaugini \cite{PF}, which therefore receive non-vanishing instanton contributions. 
 
In this paper we will only consider $\mathcal{N}=4$ SYM at the superconformal point, located at the origin of the Coulomb branch, 
and study   instanton contributions to correlation functions of protected operators, following the original analysis of  \cite{Bianchi:1998nk, Dorey:1999pd}  and  \cite{Green:2002vf,Kovacs:2003rt}. 
 We will perform most of the  computations for  $SU(2)$ gauge group, where {\it inter alia} all fermionic zero-modes are ``geometric" and the scalar propagator in the instanton background drastically simplifies. We will then discuss the modifications for $N>2$, in particular for large $N$, where they acquire a particularly simple and  elegant form.  We note that a reason to consider the large-$N$ limit is related to the fact that   the correlator/amplitude duality should only hold in the planar limit. \black The large-$N$ limit of our non-perturbative results can then be used to test the validity of this duality after the inclusion of instanton effects.
\black
Concretely,  we will consider two   classes of correlation functions: 

{\bf 1.} The ``minimal" correlation functions, where the operator insertions precisely saturate all the 16 geometric fermionic zero-modes, {\it i.e.} 
8 supersymmetric and 8 superconformal zero-modes. 
Two notable instances are the four-point correlation function of (lowest) conformal dimension $\Delta =2$ CPO's in the ${\bf 20'}$ representation of the $SU(4)$ R-symmetry  group, and the 16-point correlation function of spin-1/2 fermionic composite operators with  dimension $\Delta = 7/2$ in the ${\bf 4}$ representation of  $SU(4)$. In the following we will need a particular component of the former, namely 
\beq
{\cal G}_4 (x_1, \ldots, x_4) \ = \ \langle \Tr(Z^2)(x_1)  \Tr(\bar{Z}^2) (x_2)  \Tr(Z^2) (x_3) \Tr(\bar{Z}^2)(x_4) \rangle \,  .
\eeq

{\bf 2.}  
A second,  larger class comprises the ``non-minimal" correlation functions, where one can either  absorb non-geometric zero-modes (which are present for $N>2$) with some of the insertions, or  Wick-contract some of the insertions using the appropriate propagator in the instanton background.  
In the following we will consider non-minimal correlation functions of lowest CPO's,  specifically%
\footnote{In the following we will not explicitly indicate the spacetime dependence of each operator insertion. It will be understood that each insertion is performed at a different spacetime point.}  
\beq
{\cal G}_{2n-1} \ = \ \langle \Tr(XZ)  \Tr(\bar{Z}^2)  \cdots \Tr(Z^2) \Tr(\bar{Z}^2)  \Tr(Z\bar{X})\rangle \ , 
\eeq
and
\beq
{\cal G}_{2n} \  = \ \langle \Tr(Z^2) \Tr(\bar{Z}^2) \cdots \Tr(Z^2) \Tr(\bar{Z}^2)\rangle
\ , 
\eeq
with $n\ge 3$.  
Correlation functions of higher CPO's, dual to KK excitations of supergravity and their super-descendants, have been considered in \cite{Green:2002vf}, but we will not delve into these since they are not relevant for our present purposes. 

As we mentioned earlier,  we will begin by working with gauge group $SU(2)$  and instanton number $k=1$. In this simple case, no extra fermionic zero-modes are present in addition to the 16 geometric zero-modes. Then each elementary scalar field (be it $Z$, $\bar{Z}$, $X$ or $\bar{X}$) can either absorb two fermionic zero-modes or contract with its conjugate at a different insertion point. In fact, since we are eventually interested in the pairwise light-like limit 
$x_{i,i+1}^2 \rightarrow 0$, we will only consider contractions of fields at consecutive insertion points. 
In this limit the dominant contribution at tree level is
\be
{\cal G}^{\rm dom,tree}_{2n-1} = \langle \Tr(XZ) \Tr(\bar{Z}^2) \cdots \Tr(Z^2) \Tr(\bar{Z}^2) \Tr(Z\bar{X})\rangle = C_{2n-1} \prod_{i =1}^{2n-1}x_{i,i+1}^{-2}
\, , \ee
and
\be
{\cal G}^{\rm dom,tree}_{2n} = \langle \Tr(Z^2) \Tr(\bar{Z}^2) \cdots \Tr(Z^2) \Tr(\bar{Z}^2)\rangle
= C_{2n} \prod_{i=1}^{2n} x_{i,i+1}^{-2} \ .
\ee
We will see momentarily that  the scalar propagator in the one-instanton background produces  the same dominant contribution  in the light-like limit as the free scalar propagator.  On the other hand,  the ``uncontracted" scalars, {\it i.e.} those that are chosen to absorb the 16 fermionic zero-modes, do not produce the same kind of singularity. Actually, before integration over collective coordinates, the opposite is true --  ``uncontracted" scalars tend to produce zeros due to fermionic zero-mode repulsion, which is for instance at the heart of the vanishing of non-perturbative corrections to the anomalous dimension of the lowest Konishi operator and its super-descendants \cite{Bianchi:1999ge}.

Next, we note that in order to expose the singularity structure of the correlation functions, it is very convenient to transform to Mellin space. In particular we will   observe that corrections to the free scalar propagator due to  instantons can be expressed in terms of  bulk-to-boundary propagators  $K_\Delta(x;z)$ with $\Delta =1$, and  derivatives of the logarithm of bulk-to-boundary propagators (which may be further recast into derivatives of the bulk-to-boundary propagator $K_\Delta(x;z)$ with $\Delta = \epsilon \rightarrow 0$).
We will see that even prior to taking the consecutive light-like limit, the instanton contributions to correlation functions can be decomposed in terms of conformally invariant integrals that are naturally associated to contact interactions in AdS.  We will show that in the consecutive light-like limit, and for  $N=2$ and $k=1$, 
\beq
{{\cal G}^{\rm dom,inst}_{p}\over {\cal G}^{\rm dom,tree}_{p} } \ \rightarrow \ 0
\ .
\eeq
In view of our earlier discussion, this should be taken as circumstantial evidence for the vanishing of instanton corrections to the light-like polygonal Wilson loop -- at least at leading order in $g_{\rm YM}$ --  and, going one step further, to MHV amplitudes. We will also describe how to generalise our results to any $N$  and to generic $k$  in the large-$N$ limit.\footnote{Our results do not seem to depend in a special way on the number of insertions, in particular when moving from five to six points nothing particular happens in our instanton calculation. This is in contrast with the perturbative calculation where conformal cross-ratios and a corresponding remainder appear for six and more legs. We thank the referee for raising this issue.}

Before concluding this introduction we would like to make a comment on the precise large-$N$ limit we are considering. It is often argued that instanton corrections vanish in the 't Hooft 
limit whereby $g^2_{\rm YM} \rightarrow 0$ and $N\rightarrow \infty$ while the 't Hooft coupling $\lambda = g^2_{\rm YM}N$ is kept fixed. This  occurs because such corrections are proportional to 
$e^{- 8 \pi^2 /g^2_{\rm YM}} = e^{-8 \pi^2 N/\l}$, which vanishes as $N\to \infty$. We stress that our limit is different from the 't Hooft limit as we keep $g_{\rm YM}$ small but fixed, so that $e^{- 8\pi^2/g^2_{\rm YM}}$ is non-vanishing but exponentially suppressed compared to perturbation theory. This is the limit which is appropriate to match instanton corrections in $\cN=4$ SYM with D-instanton corrections in type IIB string theory as it was done in \cite{Bianchi:1998nk, Dorey:1999pd}. 

The plan of the paper is as follows. 
In Section \ref{propagate} we perform  a careful analysis of the scalar propagator in the one-instanton background for $SU(2)$ gauge group. This will prove crucial for our subsequent analysis. In Section \ref{fourpt-sec} we reanalyse the four-point function of lowest CPO's and show that the ratio of the one-instanton contribution to the tree-level correlator vanishes in the pairwise consecutive light-like limit. We then move on  to generalise this result to five points in Section \ref{fivept}, and to an arbitrary number of points in Section \ref{generpt}. In Section \ref{Kinst} we address the issue of (multi)-instanton contributions for $SU(N)$ gauge group and then discuss the large-$N$ limit. Section \ref{discuss} contains a discussion of several important aspects, including the possible extension of the duality with Wilson loops beyond perturbation theory, and the role of
possible higher-order corrections to our results. Finally, in Section \ref{conclus} we summarise our findings and present  our conclusions.

\section{Scalar propagator in  a one-instanton background} \label{sec-propagator}
\label{propagate}

 Our analysis focusses mostly on the non-minimal correlation functions, and a crucial ingredient for this is the scalar propagator in the one-instanton background. In this section we will carefully analyse its  structure, specifically for gauge group $SU(2)$. Our notation and some useful background on instantons  are collected in Appendix \ref{Appendix:notation} and \ref{Appendix:instanton}. 

The scalar propagator in an instanton background has been worked out in   \cite{Corrigan:1978xi} (see also \cite{Green:2002vf}). 
 For $SU(2)$ gauge group and instanton number $k=1$ its expression is given by  the remarkably simple and compact formula
\beq
G^{ab}_{i j}(x, y) = {\delta_{ij} \over 8 \pi^2 (x-y)^2 }  {\rm Tr} \Big[\sigma^a u^\dagger(x) u(y) \sigma^b u^\dagger(y) u(x) \Big],
\eeq
where the  $2 \times 1$ matrix of quaternions $u(x)$ is 
\[
 u(x) :={1 \over \rho \sqrt{\rho^2 + (x-x_0)^2 }} \left( \begin{array}{c}
 q (x-x_0)^\dagger   \\
 \rho^2
\end{array} \right).
 \]
where $x = x_\mu \sigma^\mu$ and similarly for $x_0$ and $q$, with $q^2 =\rho^2$. From this definition we find,
\beq
u^\dagger(x) u(y) = {1 \over \rho^2 \sqrt{\rho^2 + (x-x_0)^2 }\sqrt{\rho^2 + (y-x_0)^2 }} (A + B_a \sigma^a ) \, ,
\eeq
with
\bea  
A &=& \rho^2 + (x-x_0) \cdot (y-x_0) \\ \nonumber \cr
&=& {1 \over 2} \Big[ 
 (\rho^2 + (x-x_0)^2) + (\rho^2 + (y-x_0)^2) -(x-y)^2  \Big]\, , \nonumber \\ \cr \cr
B^a &=&   i \eta^a_{\mu \nu} (x-x_0)^{\mu} (y-x_0)^{\nu} 
=  {i \over 4}\eta^a_{\mu \nu} \partial^{\mu}_x  \big[ \rho^2 + (x-x_0)^2 \big] \partial^{\nu}_y  \big[ \rho^2 + (y-x_0)^2\big] 
\\ \nonumber \cr
&=& {i \over 2} \eta^a_{\mu \nu} (x-y)^{\mu} \partial^{\nu}_y  \big[ \rho^2 + (y-x_0)^2 \big] 
= {i \over 2} \eta^a_{\mu \nu} (x-y)^{\mu} \partial^{\nu}_x  \big[ \rho^2 + (x-x_0)^2 \big] \, ,  \nonumber
\eea
where $\eta^a_{\mu \nu}$ is the 't Hooft symbol \cite{hooft}. We have expressed $A$ and $B$  in several suggestive forms for later use. Combining the various terms, one obtains
\beq \label{propagator0}
G^{ab}_{i j}(x, y) = {\delta_{ij} \over 4 \pi^2 (x-y)^2 }  
{\delta^{ab}(A^2 - B^2) + 2( \delta^{ab}B^2- B^a B^b) - 2 i \epsilon^{a b c} B_c A  \over (\rho^2 + (x-x_0)^2) (\rho^2 + (y-x_0)^2)} \ .
\eeq
Using these expressions for  $A$ and $B$, the propagator can be  written as a sum of different contributions, 
\beq
\label{propagator}
{G_{ab}}_{i j}(x, y) = {\delta_{ij} \over 4 \pi^2} {\delta_{ab} \over (x-y)^2} + 
{\delta_{ij} \over 4 \pi^2} \Big[ G^{(0)}_{ab}(x,y) + G^{(1)}_{ab}(x,y) + G^{(2)}_{ab}(x,y)   \Big]  \ , 
\eeq
where we have singled out the free propagator. As we will see shortly, the remaining terms $G^{(0)}$, $G^{(1)}$  and $G^{(2)}$  admit an interesting  interpretation in terms of bulk-to-boundary propagators (or derivatives thereof) in AdS space. We now look in greater detail at the different terms. 

From the  $A^2 - B^2$ term in \eqref{propagator0}, we produce the free propagator as well as 
\beq
\label{G0}
G^{(0)}_{ab}(x,y) = -\delta_{ab}  K_{1}(x;z) K_1(y;z) \, ,
\eeq
where we have defined 
\beq\label{bbprop}
K_1(x;z) := {\rho \over \rho^2 + (x-x_0)^2} \ ,
\eeq
which is precisely a  bulk-to-boundary propagator of a scalar with conformal dimension $\Delta=1$.%
\footnote{Note that our definition \eqref{bbprop} differs from the standard bulk-to-boundary propagator by a normalisation factor ${\Gamma(\Delta) \over 2 \pi^{h} \Gamma(1+ \Delta -h)}$, where $h={d \over 2}$ and $d$ is the number of dimensions. With this standard normalisation the bulk-to-boundary propagator would be ill-defined for $\Delta = 1$ in $d=4$. } 
Here $x = (0, x^\mu)$ and $z= (\rho, x^\mu_0)$  label points on the boundary and the bulk of AdS$_5$, respectively. 
Hence, we see that  $G^{(0)}_{ab}(x,y)$ is a product of bulk-to-boundary propagators in AdS$_5$. 

Moving to the other terms, a particular piece of the term $-2 i \epsilon^{a b c} B_c A$ gives the contribution 
\beq
\label{G1}
G^{(1)}_{ab}(x,y) = 
{1 \over 2} \epsilon^{abc} {\eta_c}_{\mu \nu}  {(x-y)^{\mu} \over (x-y)^2 }
\big[ \partial_{x_{\nu}} \log \big(K_1(x;z)\big) + \partial_{y_{\nu}} \log\big(K_1(y;z)\big) \big] \, ,
\eeq
which is a sum of logarithms of bulk-to-boundary propagators with differential operators  
acting on the boundary points. Formally, we can write 
\beq
\partial_{x_{\nu}} \log\big(K_1(x;z)\big) = \lim_{\varepsilon  \to 0} \, \Gamma(\varepsilon ) \partial_{x_{\nu}} K_{\varepsilon}(x;z) \, .
\eeq
In this sense, $\partial_{x^{\mu}} \log\big( K_1(x; z) \big)$  is the derivative of a ``putative" bulk-to-boundary propagator with zero conformal dimension.  A propagator with zero conformal dimension is of course ill-defined, but the differential operator makes it well-defined. 

Finally,  from  the term $2(\delta^{ab}B^2 - B^a B^b)$ and the remaining part  of $-2 i \epsilon^{a b c} B_c A$, one obtains
\beqa \nonumber
G^{(2)}_{ab}(x,y) &=& {1 \over 4}
\Big[ - \epsilon^{abc} {\eta_{c}}_{\nu \lambda } 
+ 2 (\eta^a_{\mu \nu} \eta^b_{\kappa \lambda}-\delta^{ab}\eta^c_{\mu \nu} {\eta_c}_{\kappa \lambda} )
{(x-y)^{\mu} (x-y)^{\kappa} \over (x-y)^2} \Big] \\  \cr
&\times & \partial_{x_{\nu}} \log\big(K_1 (x; z)\big) 
 \partial_{y_{ \lambda }} \log\big(K_1(y;z)\big) 
\nonumber \cr
&= &
{1 \over 4}
\Gamma(\varepsilon )^2 \Big[ - \epsilon^{abc} {\eta_{c}}_{\nu \lambda } 
+ 2 (\eta^a_{\mu \nu} \eta^b_{\kappa \lambda}-\delta^{ab}\eta^c_{\mu \nu} {\eta_c}_{\kappa \lambda} )
{(x-y)^{\mu} (x-y)^{\kappa} \over (x-y)^2} \Big] 
\\   \cr
&\times & \partial_{x_{\nu}} \big[K_{\varepsilon} (x; z)\big]~ \partial_{y_{ \lambda }} \big[K_{\varepsilon}(y;z) \big] \, .
\label{G2}
\eeqa
In the last step we have again written the result in terms of bulk-to-boundary propagators with zero conformal dimension. 

Hence we conclude that  the scalar propagator in a one-instanton background can be written as a sum of various contributions, which have the form of  (derivatives of) bulk-to-boundary propagators in AdS$_5$ --  schematically, 
\beqa
&& G^{(0)}(x, y) \sim K_1(x; z) K_1 (y; z) \, , \\ \nonumber \cr  
&& G^{(1)}(x, y) \sim \partial_x   K_{\epsilon }(x; z)  +  \partial_y K_{\epsilon } (y; z) \, , \\ \nonumber \cr  
&& G^{(2)}(x, y) \sim \partial_x   K_{\epsilon }(x; z) \times  \partial_y K_{\epsilon } (y; z) \, .
\eeqa 
Later on we will see that, with  this decomposition of the propagator at hand,  the whole correlation function is in fact closely related to certain Witten diagrams in AdS space. 
More specifically,  the instanton moduli $\rho$ and $x_0$  will only appear through the  combination
\beq
K_{\Delta}(x_i; z) = { \rho^{\Delta} \over (\rho^2 + x_{i,0}^2)^{\Delta} }
\ , 
\eeq
which is of course the bulk-to-boundary propagator in AdS$_5$ introduced earlier. Hence a generic  correlation function of lowest CPO's is just a sum of contact terms in AdS space, which is given by gluing all the bulk-to-boundary propagators at one single integration point in the bulk, possibly with some derivative acting on  the boundary points. 
We also note that, from recent works on the Mellin integral representation of Witten diagrams\cite{Penedones:2010ue, Fitzpatrick:2011ia, Paulos:2011ie, Nandan:2011wc}, 
a contact term in AdS space is in fact the simplest Witten diagram. In Mellin space, the result of such Witten diagram is given by the compact expression
\beq
  \int^{ +i \infty }_{ - i \infty }  [d\alpha ] \prod_{i<j} (x^2_{i,j})^{-\alpha_{i,j}} \Gamma (\alpha_{i,j} )
\ , \eeq
where $\alpha_{i,i}=0$,  and the off-diagonal $\alpha_{i,j}$'s are symmetric, and subject to the constraints 
\beq
 \sum_j \alpha_{i,j} \ =  \  \Delta_i 
 \ . 
 \eeq
A  more detailed presentation of Mellin integrals, including the definition on the integration contour and the integration measure $[d\alpha ]$, can be found in Appendix \ref{appendix:mellin}.

\section{Four-point correlation function}%
\label{fourpt-sec}
We start by considering the simplest, non-trivial correlation function which receives   one-instanton corrections, namely the   four-point correlation function of lowest CPO's, 
\bea 
{\cal G}^{\rm 1-inst}_4(x_i) = \langle {\rm Tr} ( Z^2)(x_1) {\rm Tr} ( \bar{Z}^2) (x_2)
{\rm Tr} ( Z^2) (x_3)  {\rm Tr} (\bar{Z}^2) (x_4) \rangle_{k=1} \, .
\eea
This correlation function was first studied in~\cite{Bianchi:1998nk} at weak coupling for gauge group $SU(2)$ and at strong coupling using AdS/CFT.\footnote{\black See~\cite{Alday:2013cwa} for a recent analysis of this four-point correlation function in the OPE limit. \black } We will briefly review this calculation here and reformulate the result in Mellin space. 
The Mellin representation is very convenient to study consecutive light-like limit of the correlator (divided by its tree level counterpart) as we will show here for the four-point case and later for higher-point correlators.\footnote{Some of the results of this section were obtained in 2011 in collaboration with Valya Khoze and Bill Spence at the  Kavli Institute for Theoretical Physics, University of California, Santa Barbara.}

In the semiclassical approximation, the result of the correlation function in the background of an instanton is obtained by simply replacing each scalar field by the following expressions:
\bea 
\label{4ptzeromode}
Z^a \rightarrow f(x) \big( \zeta^1 \sigma^a \zeta^4 \big) \ , \ \ \bar{Z}^a \rightarrow  f(x) \big(\zeta^2 \sigma^a  \zeta^3 \big) \, ,
\eea
where 
\beq
\label{zeta}
 \zeta^A(x) \, :=\, 
  \sqrt{\rho} \left( \eta^A \, + \, {x \over \rho} \bar{\xi}^A\right), 
 \eeq
 and%
 \footnote{In the following we will often refer to $f(x)$ as the ``instanton profile". Note $f(x)$ is precisely the bulk-to-boundary propagator with conformal dimension two, $K_2(x;z)$.}
 \beq
  f(x)\, := \, {\rho^2 \over \big[\rho^2 + (x-x_0)^2\big]^2} \, .
\eeq
Here $\eta$ and $\bar{\xi}$ denote two constant spinors of opposite chirality, $A=1, \ldots, 4$ denotes an $SU(4)$ R-symmetry index and $x:=x^\mu \sigma^\mu$. In this decomposition $\eta$ and $x\, \bar{\xi}$
are the parameters of  supersymmetry and superconformal transformations, respectively. 
The expressions \eqref{4ptzeromode} 
arise from Wick-contracting a scalar field inside an operator with a scalar field from one insertion of the Yukawa action. To leading order, the fermions are replaced by their zero-mode expansion, which leads to the appearance of the supersymmetric and superconformal zero-modes in \eqref{4ptzeromode}. 

The result for ${\cal G}^{\rm 1-inst}_4(x_i)$ is then%
\footnote{We have dropped numerical factors as well as an overall factor of ${g_{\rm YM}^8 \over 32 \pi^{10}} \, e^{-{8 \pi^2 \over g_{\rm YM}^2} + i \theta_{\rm YM}}$ coming from the one-instanton measure for $N=2$.} 
\bea 
{\cal G}^{\rm 1-inst}_4(x_i) 
&=& \int\!{d \rho d^4 x_0 \over \rho^5} \, d^8 \eta \, d^8 \bar{\xi}
\  \, 
\big[  \zeta^1(x_1)\big]^2 
\big[\zeta^4(x_1)\big]^2  
\big[\zeta^2(x_2)\big]^2 
\big[\zeta^3(x_2)\big]^2 
\\  \cr
&&
\big[\zeta^1(x_3)\big]^2 
\big[\zeta^4(x_3)\big]^2
\big[\zeta^2(x_4)\big]^2 
\big[\zeta^3(x_4)\big]^2  \ 
f^2(x_1) f^2(x_2)f^2(x_3)f^2(x_4)  \, ,
\nonumber
\eea
where we have used 
\bea
\big(\zeta \sigma^a \tilde{\zeta}\big) 
\big( \zeta  \sigma_a \tilde{\zeta}\big)
\ =\ {1 \over 4} {\rm Tr} \big[ \sigma^a \tilde{\zeta}^2  \sigma_a \zeta^2 \big] 
 \ =\  {3 \over 2}\tilde{\zeta}^2   \zeta^2 \, . 
\eea
This result can be simplified considerably by noting that 
\beq
\zeta(x)^2\zeta(y)^2 = (x-y)^2 \eta^2 \bar{\xi}^2
\, , 
\eeq
thus arriving at
\beq
\label{intri}
{\cal G}^{\rm 1-inst}_4(x_i) 
\ = \ { x^4_{1,3} x^4_{2,4} } \int\!{d \rho d^4 x_0 \over \rho^5} \ 
f^2(x_1) f^2(x_2)f^2(x_3)f^2(x_4) \, .
\eeq
As observed in \cite{Bianchi:1998nk},  \eqref{intri} admits an  intriguing re-interpretation as a Witten diagram in AdS$_5$ with four bulk-to-boundary AdS propagators connecting four boundary points $x_i$'s to a common bulk point $z = (\rho, x^\mu_0)$, which is then integrated over with the standard, conformally invariant AdS$_5$ measure $\int_{AdS_5} := \int \frac{d\rho d^4 x_0}{\rho^5}$,
\bea \label{fourpt}
{\cal G}^{\rm 1-inst}_4(x_i) 
\ = \ { x^4_{1,3} x^4_{2,4} } \int_{AdS_5} K_4(x_1; z) K_4(x_2; z) K_4(x_3; z) K_4(x_4; z) \, .
\eea
Inspired by recent applications of Mellin integrals to the study of correlation functions in conformal theories, 
we can recast \eqref{fourpt} in Mellin space.
 As we will see, this is particularly useful for studying light-like limits.
We can then re-express the correlation function  as a simple, two-fold Mellin-Barnes-type contour integral%
\footnote{Here numerical factors from Mellin integrals have been dropped. We have summarised some crucial facts about Mellin integrals in Appendix \ref{appendix:mellin}.}, 
\bea
\label{mamp}
{\cal G}^{\rm 1-inst}_4(x_i) \ = \ { x^4_{1,3} x^4_{2,4} }
 \int^{ + i \infty }_{ - i \infty } [d \alpha] \prod_{1 \leq i<j \leq 4} (x^2_{i,j})^{-\alpha_{i,j}} \Gamma(\alpha_{i,j}) \, ,
\eea
where the integration variables $\alpha_{i,j}$ are constrained by 
\beq
\label{alcon}
\sum_{j=1}^4 \alpha_{i,j} =\Delta_i
\ . 
\eeq
In our particular case all the conformal dimensions are%
\footnote{Of course the operators inserted have conformal dimension $2$, as made clear in \eqref{mamp} it is compensated by the prefactor $x^4_{1,3} x^4_{2,4}$.}
$\Delta_i =4$. We will refer to the right-hand side of \eqref{mamp} as a Mellin integral. 
The Mellin integral in \eqref{mamp}  can also be understood as a product of differential operators $\prod_{i<j} \partial_{x^2_{i,j}}$ acting on the same Mellin integral but with different conformal dimensions, $\Delta'_i = 1$. The Mellin integral with $\Delta'_i = 1$ is nothing but the Mellin representation of a  four-mass box integral in four dimensions \cite{SVP}. This observation was used in \cite{Bianchi:1998nk} to obtain a closed analytic expression for the instanton correlator.

We also note that another advantage of the Mellin integral representation is to make the OPE manifest. More precisely,  poles of the integrand of the Mellin integrals correspond to operators which can appear in the OPE \cite{Fitzpatrick:2011ia}. 
The result of the one-instanton correction to the four-point correlation function shows that all poles come from the $\Gamma(\alpha_{i,j})$ functions.  In the OPE language, this means that only double-trace ``quasi-protected'' operators, dual to multi-particle supergravity states, are exchanged in the intermediate channels. Konishi and Konishi-like operators, dual to genuine string excitations in AdS, do not appear in this case. This is consistent with the OPE analysis of the correlation function ${\cal G}^{\rm 1-inst}_4(x_i)$  performed in \cite{Bianchi:1999ge}.  

Next, we turn to considering the light-like limit of \eqref{fourpt}. A major advantage of the Mellin representation is that we can take the limit directly on the integrand of the Mellin integral prior to performing the $\alpha_{i,j}$-integrations.
Thanks to the constraints \eqref{alcon} one can solve for four of the six $\alpha_{i,j}$'s in terms of the other two. One possibility is to express everything in terms of $\alpha_{3,4}$ and 
$\alpha_{1,4}$, 
\bea
\alpha_{1,2} = \alpha_{3,4}, ~~~ \alpha_{2,3} = \alpha_{1,4}, ~~~ 
\alpha_{1,3} = \alpha_{2,4}= 4 - \alpha_{1,4} - \alpha_{3,4} \, .
\eea
Taking into account these relations,  \eqref{mamp} becomes
\bea
{\cal G}^{\rm 1-inst}_4(x_i)  \nonumber
= {1 \over  x^4_{1,3} x^4_{2,4}} 
 \int^{ + i \infty }_{ - i \infty } {d \alpha_{3,4} \over 2 \pi i}  { d \alpha_{1,4}  \over 2 \pi i}
u^{-\alpha_{3,4}} v^{-\alpha_{1,4}}  
\Gamma^2(\alpha_{3,4})\Gamma^2(\alpha_{1,4}) \Gamma^2(4 - \alpha_{1,4} - \alpha_{3,4})\, , 
\eea
where we have introduced the conformal cross-ratios
\bea
u := { x^2_{1,2} x^2_{3,4} \over x^2_{1,3} x^2_{2,4} }\, , \quad v := { x^2_{2,3} x^2_{4,1} \over x^2_{1,3} x^2_{2,4} }\, . 
\eea
Since we are interested in the consecutive light-like limit, namely $x^2_{i,i+1} \rightarrow 0$ or $u, v \rightarrow 0$, 
we have to close the integration contour in the $\alpha$ variables at $Re (\alpha_{i,j}) \to - \infty$. 
To leading order in $u$ and $v$ we have
\bea
{\cal G}^{\rm 1-inst}_4(x_i) 
\sim  {1 \over x^4_{1,3} x^4_{2,4} } \log (u) \log (v) \, ,
\eea
where the logarithmic singularities arise from the residues of the integrand at $\alpha_{1,4}=\alpha_{3,4} = 0$. Dividing by the tree-level result, 
\beq
{\cal G}^{\rm tree}(x_i) \ = \ {1 \over x^2_{1,2} x^2_{2,3} x^2_{3,4} x^2_{4,1}}\, + \, \cdots \, =\,  \frac{1}{x_{1,3}^4 x_{2,4}^4 u v}\, + \, \cdots \ , 
\eeq
where the dots stand for terms that are subleading in the consecutive light-like limit, 
we find 
\bea
{ {\cal G}^{\rm 1-inst}_4(x_i)\over {\cal G}^{\rm tree}_4(x_i)} \sim  u \, v\,  \log (u)\,  \log (v) \to 0 \, , \ {\rm as} \ \ u,v \to 0 \ .
\eea 
Therefore, we conclude that the four-point correlation function (divided by the tree-level contribution) vanishes in the consecutive light-like limit. 

We end this section with three comments.

{\bf 1.} We note that any four-point correlators of generic  bilinear half-BPS scalar operators can be written in a unified form by using a  non-renormalisation property of the stress tensor correlator. The result is given in Appendix~\ref{appendix:partialNR}.

{\bf 2.} We observe that the  same correlation function  in the background of an instanton with topological charge $K>1$ and gauge group $SU(N)$ has been evaluated in \cite{Dorey:1999pd}  in the large-$N$ limit. We briefly review this result and discuss its light-like limit in Section \ref{section:largeNrev}.

{\bf 3.} 
Finally, we note that we have taken the light-like limit directly at the level of the integrand of a Mellin integral. This circumvents the problem of having to analytically continue an explicit expression to the correct analytic region (from Euclidean to Lorentzian signature). We also mention that we confirmed the correctness of the procedure followed here by taking the limit on the integrated expression of \cite{Bianchi:2007ft}.

\section{Five-point correlation function}
\label{fivept}
In this section we wish to describe one-instanton corrections to higher-point correlation functions of lowest (dimension-two) CPO's. Such quantities have not
been explicitly considered before, and belong to the broad class  of non-minimal correlators \cite{Green:2002vf}.
The reason behind this name is that, unlike the four-point case studied in the previous section, not all of the scalar fields absorb geometric (exact) fermionic zero-modes. Hence, for $SU(2)$ gauge group, the novel complication is that the remaining scalar fields have to be Wick-contracted with a scalar propagator in the background of the instanton, which was discussed in detail in Section \ref{sec-propagator}. For $N>2$, some scalar fields may absorb non-geometric (non-exact) fermionic zero-modes.  

\begin{figure}[h]
\scalebox{1}{
\centerline{\includegraphics[height=4.5cm]{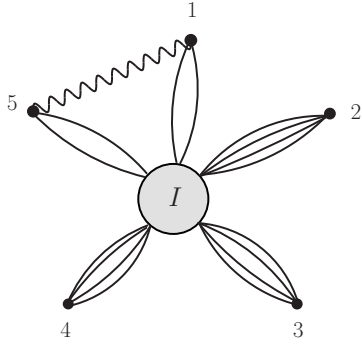} }
}
\caption{\it
A particular contribution to the five-point correlation function with scalar fields $\bar{X}(x_5)$ and $X(x_1)$ contracted. Here the wiggly line between $5$ and $1$ denote the propagator in the instanton background, curve lines indicate the number of fermionic zero modes the operators can absorb. }
 \label{fig:51contraction}
 \end{figure}

In this section we will focus on the first non-trivial example, namely that of the
five-point correlation function
\bea
{\cal G}^{\rm 1-inst}_5(x_i) = \langle {\rm Tr} (XZ)(x_1) {\rm Tr} ( \bar{Z}^2)(x_2)
{\rm Tr} (Z^2)(x_3) {\rm Tr} ( \bar{Z}^2)(x_4) {\rm Tr} ( Z\bar{X})(x_5) \rangle_{k=1} \, \ ,
\eea
which receives contributions from diagrams which involve a single Wick-contraction between $X/\bar{X}$ or $Z/\bar{Z}$ while all remaining fields absorb the fermionic zero-modes. For definiteness we focus on the diagram depicted in 
Figure \ref{fig:51contraction}.
To leading order in the semiclassical approximation we replace eight of the scalar fields by their classical solutions in the instanton background, 
\bea \label{5ptzeromode}
X^a &\rightarrow& f(x)\big( \zeta^1 \sigma^a \zeta^2 \big) \ , \qquad 
Z^a \ \rightarrow \ f(x) \big( \zeta^1 \sigma^a \zeta^4 \big) \ , 
\nonumber \\ \cr 
\bar{X}^a &\rightarrow& f(x) \big( \zeta^3 \sigma^a  \zeta^4 \big)\ ,  \qquad 
\bar{Z}^a \ \rightarrow  \ f(x) \big( \zeta^2 \sigma^a  \zeta^3 \big) 
\, ,
\eea
while the the remaining two fields are contracted with a scalar propagator in the one-instanton background.

Denoting this contribution as ${\cal G}_{51}(x_i)$, we have
\bea  
{\cal G}_{51}(x_i) &=&  \int\!{d \rho d^4 x_0 \over \rho^5} 
\ 
G^{a b}(x_5, x_1) \, 
\big(\zeta^1 \sigma^a \zeta^4\big)(x_5) \, 
\big(\zeta^1   \sigma^b \zeta^4\big)(x_1)  \, 
 \big[\zeta^2 (x_2)]^2 
 \big[\zeta^3(x_2) \big]^2
\\  \cr
&&
\big[\zeta^1(x_3)\big]^2 \big[\zeta^4(x_3)\big]^2
\big[\zeta^2(x_4)\big]^2 \big[\zeta^3(x_4)\big]^2 \ f(x_1) f^2(x_2)f^2(x_3)f^2(x_4) f(x_5)  \, .
\nonumber
\eea
To further simplify the result, we apply the Fierz identity 
\bea \label{Fierz}
\psi_{\alpha} \chi^{\beta}  = {1 \over 2} \big[ 
- {\delta_{\alpha}}^{\beta} (\psi \cdot \chi) +  {(\sigma^c)_{\alpha}}^{\beta} \big(\psi \sigma_c \chi \big)\big] 
\, ,
\eea
to find
\bea
\nonumber
&& {\cal G}_{51}(x_i) \ = \  x^4_{2,4} \int\!{d \rho d^4 x_0 \over \rho^5}\,  
G^{a b}(x_5, x_1) \,  \zeta^1(x_3)^2 \zeta^4(x_3)^2~ f(x_1) f^2(x_2)f^2(x_3)f^2(x_4) f(x_5) \\ \nonumber 
&&  {\rm Tr} 
\Big[ 
\sigma^a 
\big[  \zeta^1(x_5) \cdot \zeta^1(x_1) -  \sigma^c \zeta^1(x_5) \sigma_c \zeta^1(x_1) \big] 
\sigma^b 
\big[\zeta^4(x_1) \cdot \zeta^4(x_5) - \sigma^d \zeta^4(x_1) \sigma_d \zeta^4(x_5) \big]
 \Big]  \ .
\cr
\\
\label{5.6}
\eea
The final result can be expanded in terms of the traces with different numbers of $\sigma$'s. Doing so, we obtain
\bea
\label{G51}
{\cal G}_{51}(x_i) &=& 
{ x^4_{2,4} }  \big[ (H^2_{51; 3} + \Omega^2_{5,1; 3})  \delta^{ab}
 + 2i \epsilon^{abc} \Omega^c_{5,1;3} H_{51;3}  - 2 \Omega^a_{5,1; 3}  \Omega^b_{5,1; 3}   \big]
 \\ \nonumber \cr
&\times &  \int {d \rho d^4 x_0 \over \rho^5} G^{a b}(x_5, x_1) 
 f(x_1) f^2(x_2)f^2(x_3)f^2(x_4) f(x_5)\, , 
  \eea
where the functions $H_{51; 3}$ and $\Omega^a_{5,1; 3}$ are defined as
\bea \label{H-function}
H_{51; 3}  &=& 
\int d^2 \eta d^2 \bar{\xi} \ \ \zeta(x_5) \cdot \zeta(x_1) ~\zeta(x_3)^2 = x_{3,1} \cdot x_{3,5} \  , \\ \nonumber 
\Omega^a_{5,1; 3}  &=& \int d^2 \eta d^2 \bar{\xi} \ \ \zeta(x_5) \sigma^a \zeta(x_1) ~\zeta(x_3)^2 =  i\eta^a_{\mu \nu} x^{\mu}_{5,3} x^{\nu}_{1,3} \ .
\eea
We would like to stress the interesting fact that these two expressions are independent of the instanton moduli $\rho, x_0$, and hence in  \eqref{5.6}  we can take these terms outside the integral. 
In fact this property  holds for all diagrams, and also for $n>5$ points, as discussed in the next section.

Next, we focus on the non-trivial integral we have to perform in \eqref{G51}, 
namely 
\bea
I_{51} := \int {d \rho d^4 x_0 \over \rho^5} G^{a b}(x_5, x_1) 
 f(x_1) f^2(x_2)f^2(x_3)f^2(x_4) f(x_5) \, . 
\eea
As we discussed earlier the propagator $G^{a b}(x_5, x_1) $ can be decomposed as 
\bea  
{G_{ab}}(x_5, x_1) = {1 \over 4 \pi^2} {\delta_{ab} \over (x_5 - x_1)^2} + 
{1\over 4 \pi^2} \Big[ G^{(0)}_{ab}(x_5, x_1) + G^{(1)}_{ab}(x_5, x_1) + G^{(2)}_{ab}(x_5, x_1)   \Big] \,  , 
\eea
where $G^{(0)}, \ldots, G^{(2)}$ are defined in \eqref{G0}, \eqref{G1} and \eqref{G2}, respectively. 
Correspondingly, there will be four contributions to  $G^{a b}(x_5, x_1)$,  
\beq
\label{I-decomposition}
I_{51}\  = \ I^{\rm free}_{51} + I^{(0)}_{51} + I^{(1)}_{51} + I^{(2)}_{51} \, ,
\eeq
where each term in the sum is given by 
\beqa
I^{\rm free}_{51} &=&
c_{\rm free} I_{2, 4, 4, 4, 2} \, ,  \nonumber \\
I^{(0)}_{51} &=&
c_{0} I_{3, 4, 4, 4, 3} \, ,
\nonumber \cr
I^{(1)}_{51} &=&
c^{\mu}_{1} 
\left(\partial_{x_5^{\mu}}+ \partial_{x_1^{\mu}} \right) I_{2, 4, 4, 4, 2  }
\, , 
\nonumber \\
I^{(2)}_{51} &=&
c^{\mu \nu}_{2}
\partial_{x_5^{\mu}} \partial_{x_1^{\nu}}  I_{2 , 4, 4, 4, 2 }   \, , 
\eeqa
and we have defined 
\bea  \nonumber
I_{\Delta_1, \Delta_2, \Delta_3, \Delta_4, \Delta_5}
&=& \int_{\rm AdS_5} K_{\Delta_1}(x_1; z ) K_{\Delta_2}(x_2; z ) K_{\Delta_3}(x_3; z ) 
K_{\Delta_4}(x_4; z ) K_{\Delta_5}(x_5; z ) \\ \cr
&=&
 \int^{+ i\infty }_ {- i\infty }  [d \alpha] \prod_{i<j} (x^2_{i,j})^{-\alpha_{i,j}} \Gamma(\alpha_{i,j}) \, .
\label{ccc}
\eea
The   coefficients $c_{\rm free} $, 
$c_{0}$,  $c^{\mu}_{1}$ and  $c^{\mu \nu}_{2}$ can easily be read off from  \eqref{propagator}, \eqref{G0}, \eqref{G1} and \eqref{G2}; in the following we will not need their explicit expressions.
In the last line of \eqref{ccc} we made use of the fact, as in the four-point case, that such AdS$_5$ contact-term integrals can naturally be expressed as Mellin integrals. 

The integral in \eqref{ccc}  is a five-point contact term in AdS$_5$, and we now focus on its evaluation. 
Firstly, note that the $\alpha_{i,j}$ have to satisfy the constraints $\sum_j \alpha_{i,j} = \Delta_i $.  We find it  convenient to express the five non-consecutive $\alpha_{i, j}$ variables in terms of the five consecutive ones, $\alpha_{i, i+1}$. Doing so we get
\beq
\label{resol}
\alpha_{i,i+2} \ = \  \Delta_{i,i+2}
+ (\alpha_{i-2,i-1} - \alpha_{i,i+1} - \alpha_{i+1,i+2}) \, ,
\eeq
with
\beq
\Delta_{i,i+2} := {1 \over 2}\big(\Delta_i + \Delta_{i+1} + \Delta_{i+2} - \Delta_{i+3} -\Delta_{i+4} \big) \, . 
\eeq
Note the indices of $\Delta_{i,i+2}$ are defined by mod $n=5$, and for the case we are considering, $\Delta_{i,i+2} \geq 0$. 
Taking the constraints into account,  $I_{\Delta_1, \Delta_2, \Delta_3, \Delta_4, \Delta_5}$ becomes
\beq
\label{5ptintegral}
I_{\Delta_1, \Delta_2, \Delta_3, \Delta_4, \Delta_5} =
 \prod^5_{i=1}  (x^2_{i,i+2})^{-\Delta_{i, i+2}} 
\int^{ + i \infty }_{ - i \infty } { d \alpha_{i, i+1} \over 2 \pi i}
 (u_{i-1, i+1})^{-\alpha_{i, i+1}} \Gamma(\alpha_{i, i+1}) \Gamma(\alpha_{i, i+2}) \, , 
  \nonumber
\eeq
where the cross-ratios $u_{i,j}$ are defined as 
\beq
u_{i,j}\  := \ {x^2_{i, j+1} x^2_{j, i+1} \over x^2_{i, j} x^2_{i+1, j+1}}\, .
\eeq
This is the result  for the contribution to  the five-point correlation function where 
$X(x_1)$ and $\bar{X}(x_5)$ are contracted by a scalar propagator in the one-instanton background.  
One can similarly obtain the remaining terms where   one Wick-contracts  $Z(x_1)$ and $\bar{Z}(x_2)$, $\bar{Z}(x_2)$ and $Z(x_3)$, $Z(x_3)$ and $\bar{Z}(x_4)$, $\bar{Z}(x_4)$ and $Z(x_5)$  (along with other possible non-consecutive contractions). \black 
The corresponding contributions to the correlation function can all be expressed as a sum of contact terms in AdS$_5$ with possible derivatives acting on boundary points where two operators are contracted by the propagator. We will not list the detailed expressions here, instead we turn to the study of the behaviour of the integral \eqref{5ptintegral} under consecutive light-like limit, in order to determine how the correlation function behaves in this limit. 

In order to perform the various $\alpha_{i, i+1}$ integrations we wish to close the corresponding  contours and employ the residue theorem. Since  
$u_{i-1,i+1} \rightarrow 0$ in the consecutive light-like limit under consideration, the contours should be closed on the left, namely for $Re (\alpha_{i, i+1}) \to -\infty $. We begin by considering the $\alpha_{1,2}$ integration. Inspecting the arguments of the various $\Gamma$-functions in \eqref{5ptintegral}, and taking into account \eqref{5ptintegral}, we see that 
there are four possible contributions: one  from  $\Gamma(\alpha_{1,2})$, and the remaining  three from $\Gamma(\alpha_{1,3})$, $\Gamma(\alpha_{5,2})$ and $\Gamma(\alpha_{3,5})$. 

To begin with, we note that    $\Gamma(\alpha_{3,5})$   has simple poles for   $\alpha_{3,5} = -n$, for non-negative integer $n$. In turn this implies
\beq
\alpha_{1,2} \  = \ -\Delta_{3,5} + (\alpha_{3,4} + \alpha_{4,5}) - n \ , 
\eeq
and the corresponding leading residue  (for $n=0$) is  proportional to 
\bea \label{limit52}
u^{\Delta_{3,5} - (\alpha_{3,4} + \alpha_{4,5})}_{5,2} \rightarrow  0
\ , 
\eea
in the consecutive light-like limit, 
since $\Delta_{3,5} > 0$  and $Re(\alpha_{i, i+1}) \leq 0$. Similarly,  the contributions from $\Gamma(\alpha_{1,3})$ and $\Gamma(\alpha_{2,5})$ are also suppressed in the consecutive light-like limit. 
We conclude that, in this limit, the only surviving contribution from the  $\alpha_{1,2}$ integration arises from the poles of 
$\Gamma(\alpha_{1,2})$. The corresponding leading contribution arises from picking the pole of 
$\Gamma(\alpha_{1,2})$ at  $\alpha_{1,2}=0$, which amounts to taking the integrand, removing $\Gamma(\alpha_{1,2})$ and setting 
$\alpha_{1,2}=0$. A similar analysis can then be performed for the  $\alpha_{4,5}$  and  $\alpha_{1,5}$ integrations.

Performing the $\alpha_{1,2}, \alpha_{4,5}$ and $\alpha_{1,5}$ integrals, we are left with the following Mellin-Barnes integral, 
\bea \nonumber
I_{\Delta_1, \Delta_2, \Delta_3, \Delta_4, \Delta_5} 
\! &\sim & \! \prod^5_{i=1}  (x^2_{i, i+2})^{-\Delta_{i, i+2}} 
\int^{ + i \infty }_{ - i \infty } 
{d \alpha_{2,3} \over 2 \pi i} {d \alpha_{3,4} \over 2 \pi i}  u^{-\alpha_{2,3}}_{1, 3}  u^{-\alpha_{3,4}}_{2, 4} \Gamma(\alpha_{2,3}) \Gamma(\Delta_{4,1}+ \alpha_{2,3})  \Gamma(\alpha_{3,4}) \\ \cr
&\times &
 \Gamma( \Delta_{5,2} + \alpha_{3,4} )  \Gamma(\Delta_{1,3} - \alpha_{2,3})  
  \Gamma(\Delta_{2,4} - \alpha_{2,3} - \alpha_{3,4}) 
  \Gamma(\Delta_{3,5}  - \alpha_{3,4}) \, .
\eea
How we pick the poles in this expression will now depend on the particular term on the right-hand side of \eqref{I-decomposition} we are looking at.

To begin with, we consider $I^{(0)}_{51}$. For this term we have $\Delta_{4,1}>0$ and $\Delta_{5,2} >0$, hence the leading contribution arises from the simple poles of the  $\Gamma(\alpha_{2,3})$ and $ \Gamma(\alpha_{3,4} )$ functions. Taking the residues, we arrive at the following  simple result, 
\bea
I_{3,4,4,4,3} \sim 
 \prod^5_{i=1}  (x^2_{i, i+2})^{-\Delta_{i, i+2}} \, . 
\eea
In the consecutive light-like limit no singularity appears in the final integration. 

Next we consider the other integrals, for which   we have $\Delta_{4,1} = \Delta_{5,2} =0$. In these   cases, similarly to the four-point correlation function,  double poles are generated from  $\Gamma^2(\alpha_{2,3}) $ and $ \Gamma^2(\alpha_{3,4})$. Picking the   residues at $\alpha_{23}=0$ and $\alpha_{34} =0$, we find
\bea \nonumber
I_{\Delta_1, \Delta_2, \Delta_3, \Delta_4, \Delta_5} &\sim & 
\Gamma(\Delta_{1,3})  
  \Gamma(\Delta_{2,4} ) 
  \Gamma(\Delta_{3,5})\prod^5_{i=1}  (x^2_{i, i+2})^{-\Delta_{i, i+2}} 
\log (u_{1,3}) \log  (u_{2,4}) .
\eea
Hence, at most logarithmic  singularities appear in the  consecutive light-like limit,  just as in  the case of the four-point correlation function. The fact that we have differential operators  acting on $x_5$ and $x_1$ cannot change this behaviour, since $u_{1,3}\sim x^2_{2,3}$ and $u_{2,4} \sim x^2_{3,4}$, namely they do not contain any $x^2_{i,j}$ in the set $\{x^2_{4,5}$, $x^2_{5,1}$, $x^2_{1,2}\}$. 

In conclusion, we have found that  the five-point correlator under consideration does not contain 
the singular prefactor $\Pi_{i=1}^5 1/x^2_{i,i+1}$. Hence, the ratio of the one-instanton contribution to this correlator and the corresponding tree-level correlator vanishes in the consecutive light-like limit. 
 A similar analysis applies  to all other contractions, with the same conclusion. Summarising,  
\beq 
{{\cal G}^{\rm 1-inst}_5(x_i)\over {\cal G}^{\rm tree}_{5}(x_i)}  \to 0
\ , 
\eeq
in the consecutive light-like limit $x_{i, i+1}^2 \to 0$.

\section{General correlation functions}
\label{generpt}

We will now generalise the above analysis for four- and five-point correlation functions of lowest CPO's to the general correlators ${\cal G}_{2m}$ and ${\cal G}_{2m+1}$ defined earlier.

\subsection{Example with a particular contraction}
%
%
%
%
\begin{figure}[h]
\scalebox{1}{
\centerline{\includegraphics[height=4.5cm]{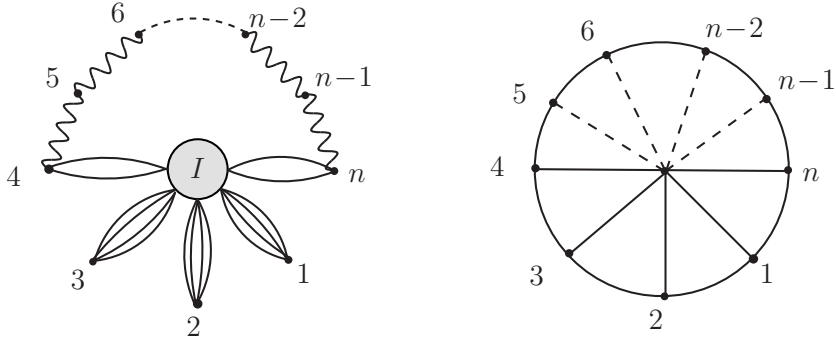} }
}
\caption{\it
The diagram on the left is the correlation function in the instanton background with connected consecutive contractions of operators at positions $x_4, \ldots, x_n$. The diagram on the right is the corresponding Witten diagram, where the dashed straight lines denote bulk-to-boundary propagators, which may or may not be present. In addition there may be differential operators acting on the boundary points $x_4, \ldots, x_n$.}
 \label{fig:npt1}
 \end{figure}
%
%
%
%
Before discussing the general case, let us begin with a special contribution to the correlation function namely the ``connected consecutive contraction" depicted in the diagram on the left of Figure \ref{fig:npt1}. To be concrete, let us consider the case where the contracted operators are ${\rm Tr}(\bar{Z}^2)(x_4), {\rm Tr}({Z}^2)(x_5), \ldots, {\rm Tr}(\bar{Z}^2)(x_{n})$, while the operators left un-contracted are ${\rm Tr}(Z^2)(x_1), {\rm Tr}(\bar{Z}^2)(x_2)$ and  ${\rm Tr}(Z^2)(x_3)$. It is straightforward to obtain the result for this particular contraction,
\bea \label{consecutive1}
\nonumber
 {\cal G}_{c} (x_i)
&=& 
\int {d\rho d^4 x_0 \over \rho^5} 
[\zeta^0(x_1)]^2 [\zeta^3(x_1)]^2 [\zeta^1(x_2)]^2 [\zeta^2(x_2)]^2 [\zeta^0(x_3)]^2 [\zeta^3(x_3)]^2
\zeta^1(x_4) \sigma^a \zeta^2(x_4) \\ \nonumber \cr
&&  \zeta^1(x_{n}) \sigma^b \zeta^2(x_{n})
f^2(x_1) f^2(x_2) f^2(x_3) f(x_4) f(x_{n}) 
\tilde{G}_{ab}(x_4, x_5, \ldots, x_{n} )  \\ \nonumber \cr
&=&   { x^4_{1,3} }  
 \big[ (H^2_{51; 3} + \Omega^2_{5,1; 3})  \delta^{ab}
 + 2i \epsilon^{abc} {\Omega_c}_{4,n;2} H_{51;3}  - 2 \Omega^a_{4,n;2}  \Omega^b_{4,n; 2}   \big]
\\  \cr
&\times & \int {d \rho d^4 x_0 \over \rho^5}  f^2(x_1) f^2(x_2) f^2(x_3) f(x_4) f(x_{n})  \tilde{G}_{ab}(x_4, x_5, \ldots, x_{n} ) \, ,
\eea
where we have defined a string of scalar propagators in the one-instanton background
\bea
\tilde{G}_{ab}(x_4, x_5, \ldots, x_{n} ) :=
G_{a b_1}(x_4, x_5) G^{b_1}_{b_2}(x_5, x_6) \ldots G^{b_{n-4}}_{b}(x_{n-1}, x_{n}) \, . 
\eea
As emphasised previously, a key observation is that the integration variables, $\rho$ and $x_0$, only appear in  $\tilde{G}_{ab}$ and in the instanton profile $f(x_i)$, which are all in the form of bulk-to-boundary propagator in AdS$_5$. 

As indicated in Figure~\ref{fig:npt1}, for each contribution to a correlation function, there is a corresponding Witten diagram. In the case we are considering, the correlation function can be schematically written as a sum of terms of the following form, 
\bea \nonumber
I_n  = \int_{AdS} K_{\Delta_1}(x_1; z) K_{\Delta_2}(x_2; z) K_{\Delta_3}(x_3; z) 
\partial^{n_4}_{4} K_{\Delta_4}(x_4; z)
\partial^{n_5}_{5} K_{\Delta_5}(x_5; z) \ldots 
\partial^{n_n}_{n} K_{\Delta_n}(x_n; z),
\eea
where $\Delta_1 = \Delta_2 =\Delta_3 =4$, while other $\Delta_i$'s depend on the choice of $n_i= 0$ or $1$. 

First of all, except for $i = 4$ and $i = n$, some $\partial^{n_i}_{i} K_{\Delta_i}(x_i; z)$'s could well be absent, since the propagator in the instanton background contains the free propagator. If they are present, we have $\Delta_i = \epsilon$ when $n_i=1$, otherwise $\Delta_i = 1$. In the case of $\Delta_i = \epsilon $, there must be a differential operator acting on the corresponding boundary point. Similarly we find that the conformal dimensions $\Delta_4$ and $\Delta_n$ can be $2$ or $3$. The details can be extracted straightforwardly from the terms $G^{(0)}, G^{(1)}$ and $G^{(2)}$ in \eqref{propagator}. For instance, one contribution 
arises from selecting the free propagator from the full propagator in the instanton background, in which case the integral reduces to 
\bea 
 \prod^{n-1}_{i=4} {1 \over x^2_{i,i+1}} \int_{AdS} K_{4}(x_1; z) K_{4}(x_2; z) K_{4}(x_3; z) 
 K_{2}(x_4; z)
 K_{2}(x_n; z) \ .
\eea
Note that this integral is nothing but one particular contribution to the five-point correlator studied earlier. Apparently the complexity increases when more  propagators in the instanton background are included. However no matter what they are, since they are contact terms in AdS space, the final integral in Mellin space is rather simple and takes the following universal form
\bea \label{consecutive}
 I_n =  \prod^{n}_{k=4} \partial^{n_k}_{k} \int^{ +i \infty }_{ -i \infty }  [d\alpha]  \prod_{i<j}(x^2_{i,j})^{-\alpha_{i,j}} \Gamma(\alpha_{i,j}) \, .
\eea 
We remind the reader that we have introduced a regulator $\epsilon$ for the propagators with zero conformal dimension, which can be set to zero after performing the appropriate derivatives.

\subsection{Arbitrary contractions}
\begin{figure}[h]
\scalebox{1}{
\centerline{\includegraphics[height=3.4cm]{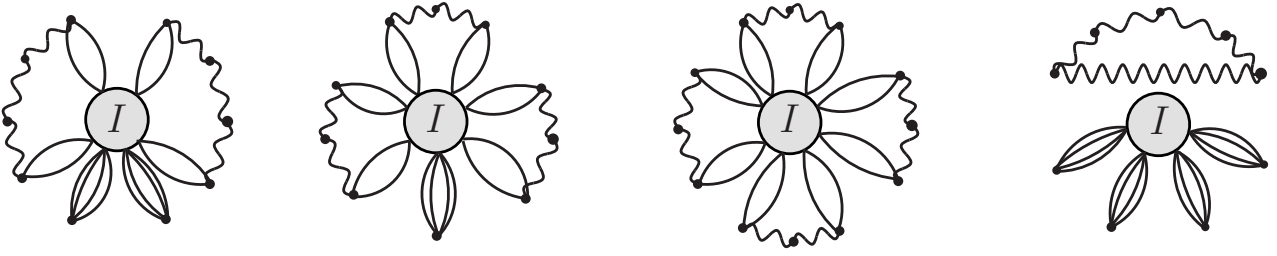} }
}
\caption{\it
Here we show the various  possible contractions for a generic higher-point correlation function, in addition to the  consecutive contraction shown in Figure~\ref{fig:npt1}.}
 \label{fig:contraction2}
 \end{figure}

The above result can be generalised to describe all possible contractions contributing to an arbitrary correlation function. In order to saturate the $16$ fermionic zero-modes, one can clearly leave at most four operators uncontracted. This case appears starting from $n=6$ points, and corresponds to having $n-4$ operators contracted independently of the remaining four operators absorbing the zero-modes. This is what one would call a ``disconnected" diagram since the propagators form a closed loop. However, because of the integration over moduli space, the diagram is not truly disconnected. Nevertheless, 
since the contractions in the ``disconnected" diagram do not mirror the contraction of the corresponding tree-level diagram, it is suppressed in the consecutive light-like limit. One can also contract more operators, as illustrated  in Figure~\ref{fig:contraction2}, where we show all other possible consecutive contractions as well as the ``disconnected" diagram. 
%
%
%
%
%
%
%
%

Still working with $SU(2)$ gauge group and $k=1$, the various contractions discussed above give rise to the following fermionic integrals:
\bea \label{newfunctions}
&& H_{pq;mn} :=\int d^2 \eta d^2 \bar{\xi} \zeta(x_p) \cdot \zeta(x_q)~  \zeta(x_m) \cdot \zeta(x_n) 
 = {1 \over 2} ( x_{p,n} \cdot x_{q,m}  + x_{p,m} \cdot x_{q,n} ) \, , \\ \nonumber \cr
&& \Omega^a_{p, q;mn} :=\int d^2 \eta d^2 \bar{\xi} \zeta(x_p) \sigma^a \zeta(x_q) ~ \zeta(x_m) \cdot \zeta(x_n)
 = i \eta^a_{\mu \nu}  ( x^{\mu}_{p,m} x^{\nu}_{q,n} - {1 \over 2} x^{\mu}_{p,q} x^{\nu}_{m,n} ) \, , \\ \nonumber \cr
&& \Omega^{ab}_{p,q;m,n} :=\int d^2 \eta d^2 \bar{\xi} \zeta(x_p) \sigma^a \zeta(x_q)~  \zeta(x_m) \sigma^b \zeta(x_n) 
={1 \over 2} \delta^{ab} x_{p,q} \cdot x_{m,n}  + {1 \over 4} \epsilon^{abc} {\eta_c}_{\mu \nu} x^{\mu}_{p,q} x^{\nu}_{m,n}
\, .
\eea
Note that $H_{pq;m}$, and  $\Omega^a_{p,q; m}$ defined in \eqref{H-function} are just special cases of $H_{pq;mn}$ and $\Omega^a_{p, q;mn}$:
\bea
H_{pq;m} = H_{pq;mm}, ~~~~\Omega^a_{p,q; m} = \Omega^a_{p, q;mm} \, .
\eea
In summary, we have found that correlation functions of CPO's can be expressed in terms of the functions appearing in \eqref{newfunctions}, the propagators in the instanton background and the instanton profile functions  $f(x_i)$'s. 

It would be a highly non-trivial task to work out precisely how the correlation functions depend on these quantities, similarly to what we did for the simple case of consecutive contractions \eqref{consecutive1}. However there is an important point that our analysis shows:  the functions  $H_{pq;mn}$ and $\Omega^a_{p, q;mn}$ are independent of the  instanton moduli $x_0$ and $\rho$, which only appear in the   propagators and instanton profiles. This general fact can also be seen in a more direct way as follows.

The functions $H_{pq;mn}$, $\Omega^a_{p, q;mn}$ and $\Omega^{ab}_{p, q;m,n}$ in \eqref{newfunctions} are particular cases of the following fermionic integral, 
\bea \nonumber
&& \int d^2 \eta d^2 \bar{\xi} \rho^2 
\Big(\eta + {\hat{x}_i - \hat{x}_0 \over \rho} \bar{\xi} \Big)^{\alpha}
\Big(\eta + {\hat{x}_j - \hat{x}_0 \over \rho} \bar{\xi} \Big)^{\beta}
\Big(\eta + {\hat{x}_k - \hat{x}_0 \over \rho} \bar{\xi} \Big)^{\gamma}
\Big(\eta + {\hat{x}_l - \hat{x}_0 \over \rho} \bar{\xi} \Big)^{\delta} \\  \cr
&& = 
\int d^2 \eta' d^2 \bar{\xi} \rho^2 
\Big(\eta' + {\hat{x}_i  \over \rho} \bar{\xi} \Big)^{\alpha}
\Big(\eta' + {\hat{x}_j  \over \rho} \bar{\xi} \Big)^{\beta}
\Big(\eta' + {\hat{x}_k  \over \rho} \bar{\xi} \Big)^{\gamma}
\Big(\eta' + {\hat{x}_l  \over \rho} \bar{\xi} \Big)^{\delta}\, , 
\eea
where in the second step we have redefined $\eta' = \eta - { (\hat{x}_0 /\rho)} \bar{\xi}$ to make the $x_0$ independence manifest. It is also easy to see that the $\rho$ dependence drops out since only terms with two $\eta^\prime$'s and two $\bar{\xi}$'s survive the fermionic integration, or equivalently by rescaling $\bar{\xi}$ integral by a factor of $\rho$. 

Because all the dependence on the moduli  $x_0$ and $\rho$ arises  from the propagators and instanton profiles,  any correlation function can be expanded in terms of the fundamental Mellin integrals 
\bea
\label{abb}
I_n\  =  \  
\left( \prod_k \partial^{n_k}_{k} \right) \int^{ +i \infty }_{ - i \infty }  [d\alpha] \prod_{i<j} (x^2_{i,j})^{-\alpha_{i,j}} \Gamma(\alpha_{i,j}) \, .
\eea
The particular conformal dimensions $\Delta_i$ and the exponents $n_k$ in  \eqref{abb}  will depend on which particular contribution one picks in the 
expression for the scalar propagator \eqref{propagator}. 
For instance, the simplest case is when only the free propagator part is taken from the scalar propagator in an instanton background.  The corresponding integral is then (see the first diagram in Figure~\ref{fig:contraction2})
\bea \label{case1}
\left( \prod_k {1 \over x^2_{k,k+1}} \right) \int_{AdS_5} K_2(x_{i-1};z) K_4(x_{i};z) K_4(x_{i+1};z) K_2(x_{i+2};z) K_2(x_j; z) K_2(x_{j+1} ;z) \, . 
\eea
With the help of Mellin integrals, we can determine explicitly how a general correlation function behaves under the consecutive light-like limit, as we will discuss in the following section. 

\subsection{Mellin integrals in the light-like limit} 
The previous discussion shows that any correlation function can be written as a sum of contact terms (and their derivatives) in AdS$_5$.  In this section we wish to study the behaviour of the Mellin integrals in the consecutive light-like limit, also in view of a possible extension of the correlator/Wilson loop duality at the nonperturbative level. To this end,  we will study the consecutive light-like limit of the following general integral, 
\bea
I_n = \prod^{n}_{k=1} \partial^{n_k}_{k} \int^{ + i \infty }_{ - i \infty }  [d\alpha] \prod_{i<j} (x^2_{i,j})^{-\alpha_{i,j}} \Gamma(\alpha_{i,j})\ .
\eea
Before solving the constraints, there are $n(n-1)/2$ $\alpha_{i,j}$ integration variables. For  $n>5$, we can always solve $n$ non-consecutive $\alpha_{k,l}$'s in terms of $n$ consecutive $\alpha_{i, i+1}$'s and  $n(n-5)/2$ non-consecutive $\alpha_{p,q}$'s. After the constrains are solved, we are left with an integral of the form
\bea \nonumber
I_n = \prod^{n}_{k=1} \partial^{n_k}_{k}  \int^{ + i \infty }_{ - i \infty } 
[d\alpha_{i, i+1}] [d\alpha_{p,q}]
\prod_{i<j} (x^2_{i,j})^{-\alpha_{i,j}} \Gamma(\alpha_{i,j})
\, , 
\eea
where we have separated the integration  into consecutive $\alpha_{i, i+1}$ and the $n(n-5)/2$ non-consecutive $\alpha_{p,q}$ variables.  

It is easy to see that the leading contributions as  $x^2_{i, i+1} \rightarrow 0 $ comes from the poles at $\alpha_{i,i+1} = 0$,  since the contours are closed at $Re (\alpha_{i,i+1}) \to -\infty$. There can be possible higher-order poles coming from multiple $\Gamma$ functions, as we have seen in the examples of four-point and five-point cases, with logarithmic divergence arising from double poles. This is the strongest  singularity that can be generated from our Mellin integral. For  higher-point contact terms, taking $\alpha_{i, i+1} \to 0$ would not lead to any higher-order poles, which means that for those cases, at leading order, we can simply remove $\Gamma(\alpha_{i, i+1})$ and set all $\alpha_{i, i+1}=0$. The result is given by
\bea
\label{511}
I_n \sim \prod^{n}_{k=1} \partial^{n_k}_{k}  
\int^{ + i \infty }_{ - i \infty } [d\alpha_{p,q}] \prod_{p,q}
(x^2_{p,q})^{-\alpha_{p,q}} \Gamma(\alpha_{p,q}) 
\prod_{k,l} (x^2_{k,l})^{-\alpha_{k,l}}  \Gamma(\alpha_{k,l}).
\eea
Note that \eqref{511} is independent of any consecutive distance $x^2_{i, i+1}$ and 
the  $\alpha_{k,l}$ are linear functions of the conformal dimensions $\Delta_i$, and of the $\alpha_{i, i+1}$ and $\alpha_{p,q}$.
 
It is now clear that we may have ${1 /x^2_{i, i+1} }$  singularities from the propagators contracting operators as well as logarithmic singularities from lower-point (four- and five-point) contact terms in AdS$_5$. Essentially any scalar fields which are not contracted by  propagators do not generate the required ${1 / x^2_{i, i+1} }$ singularity.  However those scalar fields are necessary in order to soak up the sixteen fermionic zero-modes. In conclusion, for the correlation function under investigation, 
there is no way to generate the required ${1/x^2_{i, i+1} }$ singularities for all $i$, and hence, the ratio of the instanton correlation function with its tree-level counterpart vanishes in the consecutive light-like limit {\it viz.}
\beq 
{{\cal G}^{\rm 1-inst}_n(x_i)\over {\cal G}^{\rm tree}_{n}(x_i)}  \to 0
\ , 
\eeq
in the limit $x_{i, i+1}^2 \to 0$.

\section{Charge-$k$ instantons with $SU(N)$ gauge group}
\label{Kinst}

In this section we consider correlation functions of lowest CPO's in a multi-instanton background with $SU(N)$ gauge group. Multi-instanton calculus is notoriously difficult for generic correlation functions. However, in order to make contact with the expectation value of light-like Wilson loops and, potentially, MHV amplitudes, we are mainly interested in the  consecutive light-like limit of the correlation function. 

\subsection{Four-point correlation function} \label{section:largeNrev}

Let us begin with the generalisation of the results obtained for $SU(2)$ to 
$SU(N)$ gauge group while initially focusing on the one-instanton sector. Besides the position $x_0$ and size $\rho$, in order to fully characterise the bosonic coordinates (``moduli") of the instanton, we now also need an additional set of variables, $w_{u \dot{\alpha} }$ and $\bar{w}^{\dot{\alpha} u}$ (where $u=1, \ldots N$ is a colour index and $\dot{\alpha}=1,2$ is a spinor index), to parameterise the $SU(2)$ colour orientations  and the embedding of the instanton solution in $SU(N)$. As far as fermionic coordinates are concerned, in addition to $16$ geometric zero-modes, $\eta^A$ and $\bar{\xi}^A$, as in the case of $SU(2)$, we now also have $8N-16$ extra non-geometric zero-modes, which can be expressed in terms of $8N$ variables $\nu^A_u$ and $\bar{\nu}^{Au}$ constrained by the relations
\bea
\bar{w}^{\dot{\alpha} u} \nu^A_u =0\, , \qquad w_{u \dot{\alpha} } \bar{\nu}^{Au}=0 \ .
\eea
Non-geometric zero-modes, which are present only for $N>2$,  are different in nature compared to the geometric ones, in particular they are lifted and appear explicitly in the instanton action \cite{Dorey:1999pd},
\bea
S_{\rm inst} = -2 \pi i \tau + S_{4F} \ =   \  
-2 \pi i \tau + {\pi^2 \over 2 g^2_{\rm YM} \rho^2 } \epsilon_{ABCD} \mathcal{F}^{AB}\mathcal{F}^{CD}\,    ,
\eea
where 
\bea
\tau = {4 \pi i \over g^2_{\rm YM}} + {\theta \over 2 \pi} \, .
\eea
$\mathcal{F}^{AB}$, the source of lifting non-geometric zero-modes $\nu^A_u$ and $\bar{\nu}^{Au}$, is defined as
\bea
\mathcal{F}^{AB} = {1 \over 2 \sqrt{2} } ( \bar{\nu}^{Au} \nu^B_u - \bar{\nu}^{Bu} \nu^A_u )
:= {1 \over 2 \sqrt{2} } (\bar{\nu} \nu )_{6}  \, ,
\eea
where in the last term we have indicated explicitly  that $\mathcal{F}^{AB}$ is in the ${\bf 6}$ representation of $SU(4)$. 

Hence, in the case of $SU(N)$ the classical profile for an operator $\mathcal{O}$ in the multi-instanton background generically contains both geometric zero-modes (encoded into the combination $\zeta(x)$ introduced in \eqref{zeta}) 
and non-geometric zero-modes $\nu$ and $\bar{\nu}$. For the operator we are interested in, the schematic zero-mode structure  is given by \cite{Green:2002vf}
\bea \label{UNsolution}
\mathcal{O}_{20'} \sim (f^4) (\zeta \zeta \zeta \zeta) + (f^3) \big[ \zeta \zeta ( \bar{\nu} \nu)_{10} \big]
+ (f^2) \big[ (\bar{\nu} \nu )_{10} (\bar{\nu} \nu )_{10}  \big] \, ,
\eea
where only the symmetric combination $(\nu \bar{\nu})_{10} = \bar{\nu}^{A u} \nu^B_u + \bar{\nu}^{B u} \nu^A_u$ appears. This plays an important role in the large-$N$ counting. As pointed out in \cite{Green:2002vf}, in the large-$N$ limit $(\bar{\nu} \nu)_{10}$ only contributes a factor of $g_{\rm YM}$,  whereas $(\bar{\nu} \nu)_{6}$ contributes a factor of $g_{\rm YM}N^{1/2}$. 

With this set-up, an $n$-point correlation function of operators $\mathcal{O}_i$ in the semiclassical approximation takes the form
\bea
\langle \mathcal{O}_1  \mathcal{O}_2 \ldots \mathcal{O}_n  \rangle
\ = \ 
\int d \mu_{\rm phys} \, e^{-S_{\rm inst}} 
{\mathcal{O}_1} {\mathcal{O}_2} \ldots {\mathcal{O}_n}\, , 
\eea
where some operators $\mathcal{O}_i$  have to be replaced by  their classical instanton profile ${\mathcal{O}^{(0)}_i}$ in order to saturate fermionic zero-modes, while all the others may be contracted using the propagator in the background of instantons as in the case of $SU(2)$. 
The physical measure is given by
\bea 
&& \int d \mu_{\rm phys} \, e^{-S_{\rm inst}} = 
\frac{\pi^{-4N}g^{4N} e^{2\pi i\tau}}{(N-1)!(N-2)!} \times 
\label{physmeasure} \\ 
&& \times \int {d\rho \over \rho^5}\,d^4x_0 \, \rho^{4(N-2)} \prod_{A=1}^4 
d^2\eta^A d^2\bar\xi^A \, d^{N-2}\nu^A d^{N-2}\bar\nu^A 
\, e^{-S_{4F}} \, . \nonumber
\eea 
The generalisation of the one-instanton analysis to the $k$-instanton was carried out in  \cite{Dorey:1999pd}
and requires the full machinery of the ADHM construction \cite{Atiyah:1978ri}. For arbitrary $n$, even in $\mathcal{N}=4$ SYM the general results in the $k$-instanton background are not particularly enlightening. However it has been shown  in \cite{Dorey:1999pd} that
in the large-$N$ limit and for the cases of the minimal correlation functions (such as four-point scalar correlation function we considered previously), the profile of the operator in the $k$-instanton background is simply proportional to the one-instanton expression \cite{Dorey:1999pd}, 
\bea
\left.\mathcal{O}^{(0)} \right|_{k{\rm-inst}}  = k \left. \mathcal{O}^{(0)} \right|_{k=1}\, , 
\eea 
since, roughly speaking, at large $N$ the dominant contribution comes from $k$ one-instanton configurations residing in $k$ mutually commuting $SU(2)$ factors inside $SU(N)$, all with the same size and centre.
Consequently, in the large-$N$ limit, the space-time dependence of a four-point correlation function in the $k$-instanton background is fully captured by the one-instanton result, namely
\beqa
&& 
\langle {\rm Tr}( Z^2)(x_1){\rm Tr}( \bar{Z}^2)(x_2)
{\rm Tr}( Z^2)(x_3) {\rm Tr}( \bar{Z}^2)(x_4) \rangle_{k} 
\cr \cr
 &\sim & 
\langle {\rm Tr}( Z^2)(x_1) {\rm Tr}(\bar{Z}^2)(x_2)
{\rm Tr}( Z^2)(x_3) {\rm Tr}( \bar{Z}^2)(x_4) \rangle_{k=1} \, ,
\eeqa
which then also vanishes in the light-like limit after dividing by the corresponding tree-level correlation function.

\subsection{General correlation functions}

As we have seen in earlier sections, in a generic correlation function the $16$ geometric zero-modes have to be absorbed by some of the scalar fields in the correlator. This is because the $\cN=4$ SYM action evaluated on an instanton background does not depend on these geometric zero-modes and hence cannot absorb them.
 As for the remaining   $8 Nk-16$ non-geometric zero-modes, they can be
saturated in two ways: either by explicit operator insertions or by the $\cN=4$ SYM action. 

As we mentioned earlier the $\mathcal{O}_{20^\prime}$ operator only depends on non-geometric zero-modes through $(\bar{\nu} \nu)_{10}$. Hence, operator insertions do not give rise to any factor of $N$ in the large-$N$ limit, and are therefore  equally important for the large-$N$ counting. However, it is easy to check that the first possibility does not give rise to any $1/(x-y)^2$-like singularity in the consecutive light-like limit we are considering. Hence, all non-geometric zero-modes have to be absorbed by the $\cN=4$ SYM action. As a consequence, all the remaining scalar fields have to be contracted using the propagator $G(x, y)$, which contains a ${1/(x-y)^2}$ singularity. 

The scalar field propagator in a multi-instanton background $G(x, y)$ for arbitrary $N$ and $k$ has the form
\beq
\label{SUNprop}
G^{ab}(x, y) \ = \ {{\rm Tr}\left( T^a U^\dagger(x) U(y) T^{b}  U^\dagger(y) U(x) \right)  \over 2 \pi^2 (x-y)^2} \, + \, \cdots
\ , 
\eeq
where the dots stand for terms without $1/(x-y)^2$ singularity, \black and  $U(x)$ is 
a complex $(2 k+N) \times N$ ADHM matrix defined {\it e.g.} in \cite{Dorey:1999pd,Bianchi:2007ft,Amati:1988ft,Dorey:2002ik}. It is
the $SU(N)$ generalisation of the matrix $u(x)$ introduced in Section \ref{sec-propagator}.   \black The contractions are not different from the case of $SU(2)$ and $k=1$ we have discussed in detail. As shown in Figure \ref{fig:npt1} and Figure \ref{fig:contraction2}, there are always scalar fields left over to absorb the $16$ geometric zero-modes. In the semiclassical approximation, these scalar fields will again be replaced by their classical solutions in the background of a charge-$k$ instanton. Two scalar fields at points, say $x$ and $y$, as well as propagators possibly connecting them from both sides certainly do not generate any singularity by taking the light-like limit $(x-y)^2 \to 0$. 

As we learned from the case of $SU(2)$ and $k=1$, we do not expect that $1 /(x-y)^2$-like singularities will be generated by the integration over $\rho$ and $x_0$, although we are not able to show this explicitly without knowing the detailed structure of the propagator $G(x, y)$ and of the profiles of the scalar fields. However, one can always consider the correlation function as a series expansion in $(x-y)^{\mu}$, as we will show in the following. In the case of the five-point correlator, the first term of the expansion is simply given by free propagators multiplying a four-point correlator, which is ultimately related to the four-point correlator for $SU(2)$ gauge group and $k=1$ as we discussed in 
Section \ref{section:largeNrev}. This  confirms our general expectation that the ratio of a correlator evaluated in an instanton background for any $N$ and $k$, and the corresponding tree-level correlator, vanishes in the consecutive light-like limit, just as in the case of $SU(2)$ and $k=1$. 

For concreteness, we consider the contraction shown in Figure \ref{fig:51contraction}, which is explicitly given by 
\bea\label{exam}
\int d \mu_{\rm phys}\, G^{ab}(x_5, x_1) 
Z^a_{0}(x_5) Z^b_{0}(x_1) {\rm Tr}(\bar{Z}_0^2)(x_2)  {\rm Tr}(Z_0^2)(x_3)  {\rm Tr}(\bar{Z}_0^2)(x_4) \, .
\eea   
This expression can be expanded as a series in $(x_1-x_5)^{\mu}$. The numerator in the propagator becomes
\bea \label{expansion}
{\rm Tr}\left( T^a U^\dagger(x) U(y) T^{b}  U^\dagger(y) U(x) \right)
= {\rm Tr}\left( T^a U^\dagger(x) U(x) T^{b}  U^\dagger(x) U(x) \right) +  \cdots \, ,
\eea
where the dots stand for higher-order terms in $(x-y)^{\mu}$. The leading term in this expansion is simply
\bea
{\rm Tr}\left( T^a U^\dagger(x) U(x) T^{b}  U^\dagger(x) U(x) \right) ={\delta^{ab} \over 2}  \, ,
\eea 
since $U^\dagger(x) U(x) = 1$, which gives the free propagator. 

The corresponding contribution is simply
\bea
\label{1111}
{1 \over 4 \pi^2 x^2_{51}} \int d \mu_{\rm phys}\,  
{\rm Tr}(Z^2_{0} )(x_1)  {\rm Tr}(\bar{Z}_0^2)(x_2)  {\rm Tr}(Z_0^2)(x_3)  {\rm Tr}(\bar{Z}_0^2)(x_4) \, ,
\eea
where we have used that $Z^a_0 (x_5) Z^b_0 (x_1) G^{ab} (x_5, x_1) \to \Tr (Z_0^2)(x_1)$ as $x_1\to x_5$.  
Note that \eqref{1111} is  precisely the four-point correlation function which, as discussed in the previous section,  is simply proportional to the same correlator for gauge group $SU(2)$ and $k=1$. 

Note that for some of the contractions, for instance that of $\bar{Z}(x_2)$ with $Z(x_3)$, a new type of operator insertion appears, namely ${\rm Tr} ( Z \bar{Z})$,  which is no longer half BPS. However, we note that  operators such as ${\rm Tr} ( Z \bar{Z})$ may be decomposed into two half-BPS operators plus the Konishi operator, 
\bea
{\rm Tr} ( Z \bar{Z}) \ = \ 
{1 \over 3} {\rm Tr} ( Z \bar{Z} - X \bar{X}) + {1 \over 3} {\rm Tr} ( Z \bar{Z} - Y \bar{Y})
+ {1 \over 3} {\rm Tr} (X \bar{X} + Y \bar{Y} + Z \bar{Z} ) \, . 
\eea
However it is straightforward to show that the Konishi operator vanishes when the scalar fields get replaced by the zero-modes \cite{Bianchi:2001cm},
\bea
{1 \over 3} {\rm Tr} (X_{0} \bar{X}_0 + Y_0 \bar{Y}_0 + Z_0 \bar{Z}_0 ) = 0 \, .
\eea 
In other words, the Konishi operator does not contribute to the correlation function in the instanton background. \black

Summarising, we have learnt that every propagator in an instanton background has a leading term with the appropriate  $1/x^2_{i,i+1}$ singularity, which would survive the division of the instanton correlator by the corresponding tree-level correlator. However, in order to absorb the $16$ geometric fermionic zero-modes, there must also be some scalar fields that are not contracted  and that  are replaced by the appropriate classical solution in the instanton background. Precisely the insertion of such classical solutions fails to produce the required $1/x^2_{i,i+1}$ singularities. We confirmed this by considering the simplest case, namely the five-point correlator.  Expanding in powers of  $(x-y)^{\mu}$ we have found that the non-minimal correlator reduces to a minimal one, which is in turn proportional to the correlator in the background of one 
$SU(2)$ instanton.

\section{Discussion}
\label{discuss}
In this section we would like to address a number of conceptual and practical issues. To begin with, we will discuss the possibility of a non-perturbative extension of the duality between correlation functions and polygonal Wilson loops in the light-like limit.  Because of the amplitude/Wilson loop duality, it is then interesting to consider instanton corrections to amplitudes, as computed directly by LSZ reduction.  We will then move on to consider some important practical issues related to the role of higher-order perturbative corrections to the instanton result, in particular discussing whether these corrections may   change our conclusions significantly.

\subsection{On a possible non-perturbative duality with Wilson loops in the  light-like limit}

It was shown in \cite{Alday:2010zy} that in the consecutive light-like limit $x_{i, i+1}^2 \to 0$, the correlation function of $n$ 
CPO's is dual to a polygonal Wilson loop with $n$ light-like edges, $\cC_n$. This was supported by calculations done in perturbation theory up to two loops. Remarkably,  the perturbative duality discovered in \cite{Alday:2010zy} holds even outside the large-$N$ limit, the precise statement being that \black
\beq
\lim_{x_{i, i+1}^2 \to 0} { { \cal G}^{(n)} \over { \cal G}^{(n)}_{\rm tree} }\ = \  \cW_{\rm adjoint} [ \cC_n]
\ , 
\eeq
where 
\beq
\label{wil}
\cW_R[ \cC]  \ := \ {1 \over d_R } \left\langle  {\rm Tr}_R \, \cP \exp \left[ i g\oint_{\cC} \! d\tau  A_\mu (x(\tau )) \dot{x}^{\mu} (\tau )  \right] \right\rangle
\ .
\eeq  
Here $d_R $ is the dimension of the representation $R$ and $x^\mu (\tau)$ parameterises the loop $\cC$.  \black

Since this duality holds to all orders in perturbation theory for any $N$,  it is tempting to propose that it should also survive non-perturbative, instanton corrections, and in this paper we have started addressing  this issue by considering the pairwise light-like limit of the instanton contribution to correlation functions, limiting ourselves mostly to gauge group $SU(2)$, with a brief excursus to $SU(N)$ in the large-$N$ limit.

Finding a convincing general proof of this conjecture is a hard task at present. The technical reason behind this is the treatment of the fermionic zero-modes -- particularly on the Wilson loop side. The four-point correlation function/four-edged Wilson loop contains all the main technical difficulties and we will briefly discuss this now. 

To begin with, we observe that the instanton contribution to \eqref{wil} vanishes to lowest order in the coupling constant since fermionic zero-modes would not be saturated. Following the procedure of \cite{Bianchi:1998nk, Dorey:1999pd}, one can generate  solutions to the equations of motion in the background of an instanton by applying iteratively supersymmetry transformations to the starting configuration $A = A_{\rm inst}$, with all the other fields being equal to zero.%
\footnote{This procedure is briefly reviewed in Section \ref{bbbe} below.}
At the fourth iteration, the gauge field is modified by the addition of a quartic term $A^{(4)}$ in the fermionic zero-modes, $A \to A_{\rm inst} + A^{(4)}$. 
For instanton number $k=1$ and gauge group $SU(2)$ there are 16 fermionic zero-modes, and one would have to compute the expectation value (schematically)
\beq
\left\langle 
\int\! A^{(4)} \int\! A^{(4)} \int\! A^{(4)} \int\! A^{(4)} 
+ 
\int\! A^{(4)} \int\! A^{(4)} \int\! A^{(8)}  \ + \ \cdots 
\right\rangle
\ , 
\eeq
where dots represent other possible configurations which also saturate $16$ fermionic zero-modes. It would be a rather difficult task to evaluate explicitly the above expectation value, since the form of $A^{(4)}$ and beyond is currently not known. On the other hand, assuming the correlation function/Wilson loop duality allows us to make the prediction that the instanton correction to the expectation value of a Wilson loop with light-like polygonal contour should vanish, at least at the lowest order we are considering. 

A word of caution should be added at this point, as the very definition of a light-like Wilson loop seems to clash with the Euclidean signature needed in order to define instantons, where the condition $x_{i, i+1}^2 \to 0$ implies that $x_i\to x_{i+1}$. There is however an operative way to define the limit, namely we specify the light-like contour only after analytically continuing the Green function back to Minkowski space, as we do for the correlation functions. Focusing 
on the terms discussed above, we perform the computation of
$ \langle   A^{(4)} (x_a)  A^{(4)} (x_b) A^{(4)} (x_c) A^{(4)} (x_d) \rangle $
in Euclidean space for arbitrary $x_a, \ldots, x_d$, then continue the result back to Minkowski space, and finally perform the integrations over $x_a (\tau_a), \ldots, x_d (\tau_d)$ along  the prescribed light-like  contour where, for instance, $x_a (\tau) = x_a + \tau(x_b-x_a)$, with $(x_a - x_b)^2 =0$, with $\tau\in(0,1)$.

\subsection{Direct calculation of amplitudes via LSZ reduction}
\label{direct}

In principle, the vanishing of instanton corrections to the MHV amplitudes can be checked straightforwardly by a direct computation making use of LSZ reduction formula. To apply LSZ reduction in order to calculate instanton corrections to scattering amplitudes in the MHV sector, we first need to compute a correlation function of fundamental fields, for instance 
\bea \label{mhvLSZ}
\langle A^{(4)}_1 A^{(4)}_2 A^{(0)}_3 A^{(0)}_4 \cdots A^{(0)}_n  \rangle \, .
\eea
Then, after continuing back to Minkowski signature and amputating the external legs, the above correlation function would be related to an $n$-point scattering amplitude with two negative gluons and $n-2$ positive gluons, $\mathcal{A}(1^-, 2^-, 3^+, 4^+, \ldots, n^+)$. Clearly the correlator \eqref{mhvLSZ}  can only absorb $8$ geometric fermionic zero-modes, which leads to the conclusion that instanton corrections to MHV amplitudes in $\mathcal{N}=4$ SYM vanish. However, the application of the LSZ reduction formula to instanton correlation functions is rather subtle due to the presence of infrared divergences, and will be considered in another publication \cite{LSZ}.

\subsection{Higher-order corrections to the super-instanton }
\label{bbbe}
So far we have only considered the instanton contributions to correlation functions of CPO's at lowest order in the Yang-Mills coupling $g_{\rm YM}$, and  we will now discuss possible higher-order contributions.

A field configuration with just the gauge instanton turned on and all other fields being zero is a solution to the equations of motion, but there exist more general solutions where other fields are turned on as well. In particular,
one can build the corrections to the various field components via an iterative procedure in the number of Grassmann variables. This procedure, which we have used in this paper, produces a solution of the schematic form
\beqa
\label{iter}
A &=& A^{(0)} \, + \, A^{(4)} \, + \cdots \, , 
\nonumber \\
\lambda &=& \lambda^{(1)} \, + \, \lambda^{(5)} \, + \, \cdots\, , 
\nonumber \\
\phi &=& \phi^{(2)} \, + \, \phi^{(6)} \, + \, \cdots \, , 
\nonumber \\
\bar\lambda &=& \bar\lambda^{(3)} \, + \, \bar\lambda^{(7)} \, + \, \cdots \, , 
\eeqa
where $A^{(0)}$ is the instanton, $\lambda^{(1)} $ are the fermionic zero-modes and the fields $A^{(n)}$, $\lambda^{(n)}$, $\bar\lambda^{(n)}$, $\phi^{(n)}$ with $n>1$ contain $n$ fermionic zero-modes. These higher-order terms are created iteratively due to the fact that lower-order terms produce source terms in the equations of motion.
An important example heavily used in our work is $\phi^{(2)}$ whose explicit form can be found in \eqref{4ptzeromode} and which arises from source terms that are bilinear in $\lambda^{(1)}$. 
The pattern in \eqref{iter} shows that a new term appears for each field after every four iterations and this extra term is suppressed by an additional power of $g_{\rm YM}^2$. 

One important question is whether the various corrections to the lowest-order super-instanton $\{A^{(0)}, \lambda^{(1)}, \phi^{(2)},  \bar\lambda^{(3)} \}$
are relevant for the calculation of correlation functions. This issue was studied in \cite{Green:2002vf,Kovacs:2003rt}, with the following basic conclusion: higher-order corrections to fundamental fields are relevant and non-vanishing, although suppressed by extra powers of $g_{\rm YM}^2$. Once combined into gauge-invariant composite operators, only terms with a particular structure in the fermionic zero-modes survive. In these terms, the supersymmetric and superconformal zero-modes always appear combined into the quantity $\zeta (x)$ introduced  earlier  in \eqref{zeta}, unlike in the fundamental fields. 
The non-geometric zero-modes can appear in these operators only in pairs and the maximum number of such pairs is $4N-8$. However, in order to explore this question further would require the construction of the higher-order terms appearing in \eqref{iter}.

\subsection{Instanton contributions to anomalous dimensions}

In \cite{Kovacs:2003rt}, the question as to whether composite operators  $\cO$ may receive instanton corrections to their anomalous dimensions was re-examined, by carefully studying the zero-mode structure of the two-point function $\langle \cO (x) \cO^\dagger (0)\rangle$. The main result of that paper is that only terms in the expansion of $\cO$ in Grassmann coordinates which have the structure $ \prod_{A=1}^4 \zeta^{A} (x)^2$ contribute to the correlation function, and hence to the anomalous dimension. The remaining zero-modes can appear in any way. 

In \cite{Kovacs:2003rt} a number of single- and multi-trace operators were studied, with the result that only multi-trace operators can receive instanton corrections to their anomalous dimensions. This includes both multi-trace operators built out of single-trace operators belonging to half-BPS multiplets, as well as long multiplets. Note in particular that the Konishi operator  has vanishing instanton corrections, as already discussed in \cite{Bianchi:2001cm},  at least to leading order in $g^2_{\rm YM}$. It is an open question to establish whether higher-order corrections in $g_{\rm YM}$, notably related to the use of the iterated solution $\phi^{(6)}$, may change this conclusion.%
\footnote{We thank Stefano Kovacs for discussions on this point.}

Finally, we note that  such considerations would extend also to twist-two operators and imply the absence of instanton corrections to the dimensions of such operators, \black  in agreement with 
  \cite{Basso:2007wd} which found non-perturbative corrections to the cusp anomalous dimension at strong coupling which however cannot be explained in terms of instantons, as already mentioned in the Introduction. The absence of instanton corrections to the cusp anomalous dimension in turn \black lends some indirect support to the conjecture that instanton corrections to MHV amplitudes and light-like Wilson loops in $\cN=4$ SYM  vanish, since the cusp anomalous dimension governs the universal infrared/short-distance divergences of scattering amplitudes/Wilson loops.

\section{Conclusions}
\label{conclus}

In this paper we have considered instanton corrections to the large class of non-minimal correlation functions of lowest CPO's, and we have further discussed the consecutive light-like limit of such correlators. This allows us to make the first attempt at addressing general questions regarding instanton contributions to the expectation values of light-like Wilson loops as well as to scattering amplitudes in $\mathcal{N}=4$ SYM.

By identifying an intriguing relation between such correlation functions and contact terms in AdS space, we were able to express those correlators, for the case of $SU(2)$ gauge group and instanton number $k=1$, 
 in terms of a simple universal Mellin integral, 
\bea  
 \int^{ +i \infty }_{ - i \infty }  [d\alpha] \prod_{i<j} (x^2_{i,j})^{-\alpha_{i,j}} \Gamma(\alpha_{i,j}) \, ,
\eea
possibly with some derivatives acting on boundary points. Armed with this  representation of the correlators, we have shown generally that the ratios of instanton corrections to correlation functions of lowest CPO's with the corresponding tree-level correlators vanish in the consecutive light-like limit, at least at the lowest order in the Yang-Mills coupling $g_{\rm YM}$. Assuming the validity of the correlator/Wilson loop duality, this in turn implies the vanishing of instanton contributions to the expectation value of the light-like, polygonal  Wilson loops. 

To make contact with MHV amplitudes, which are dual to correlation functions in the light-like limit only at large $N$, we further discussed the possible extension of our results to $SU(N)$ gauge group with arbitrary instanton number $k$ in the large-$N$ limit. A key lesson we extracted from the $SU(2)$, $k=1$ instanton case is that in order to absorb the $16$ geometric fermionic zero-modes, one needs to replace $8$ scalars by their classical solutions in the instanton background. However, the integration of the resulting integrand over the instanton moduli space does not give rise to $1/(x-y)^2$-like singularities, unlike conventional Wick contractions. Therefore, the ratio of the instanton correction of the correlator and the corresponding tree-level correlator vanishes in the consecutive light-like limit.

Furthermore, using results of \cite{Dorey:1999pd}, we showed that the four-point correlator with multi-instanton corrections in the large-$N$ limit has exactly the same space-time dependence as the correlator in the one-instanton background. 
We believe that this and further observations made in Section \ref{Kinst} provide strong evidence
that the ratio of correlators in $SU(N)$ multi-instanton backgrounds and their tree-level counterparts
vanish in the consecutive light-like limit, at least at leading order in $g_{\rm YM}$. This result would lead us to conclude that MHV amplitudes in $\mathcal{N}=4$ SYM do not receive instanton corrections if the correlator/amplitude duality holds beyond the perturbative level \black ({\it i.e.} in
the planar limit and in each instanton sector separately).  \black

It would be of great interest to understand if the existing perturbative proof of the correlator/Wilson loop duality \cite{Alday:2010zy}, which does not rely on any large-$N$ limit, can be extended to include non-perturbative/instanton effects.
As we mentioned, this seems to be a difficult task and is outside the scope of this paper. However we have emphasised that the vanishing of instanton corrections to MHV amplitudes and light-like Wilson loops is indirectly supported by the known fact that Konishi-like operators do not receive instanton corrections. The same should also be true for higher-spin operators associated to string states. Those facts suggest that there are no instanton corrections to the cusp anomalous dimension, which enters the results of light-like Wilson loops and scattering amplitudes.
In a separate publication we will report on the direct calculation of the instanton contribution to scattering amplitudes in $\cN=4$ SYM using the LSZ reduction \cite{LSZ}.

\section*{Acknowledgements}

It is a pleasure to thank  Valya Khoze and Bill Spence  for early collaboration on the light-like limit of correlation functions, and Marco Bochicchio, Lance Dixon, Francesco Fucito, Michael Green, Johannes Henn, Paul Heslop, Gregory Korchemsky, 
Stefano Kovacs, Radu Roiban, Giancarlo Rossi, Yassen Stanev, Emery Sokatchev and Gabriele Veneziano for stimulating discussions. CW would like to thank Joe Polchinski for discussions at early stages of the project.
We would like to thank the Department of Mathematical Sciences, Durham  University, and Herbert Gangl and Paul Heslop for their warm hospitality during the 
LMS/EPSRC Durham Symposium on ``Polylogarithms as a Bridge between Particle Physics and Number Theory". AB, GT and CW would also like thank CERN for their warm hospitality during the workshop ``Amplitudes, Strings and Branes". 
While at Queen Mary, the work of MB was supported by a  Leverhulme Visiting Professorship. The work of AB, GT and CW 
was supported by the STFC Grant ST/J000469/1,  
``String theory, gauge theory \& duality''. This work was also partially supported by the MIUR-PRIN contract 2009-KHZKRX ``Symmetries of the Universe and of the Fundamental Interactions" and the ERC Advanced Grant n.~226455 ``Superfields". 

\newpage

\appendix 

\section{$\mathbf{\sigma}$-matrix and spinor conventions}
\label{Appendix:notation}
Here we summarise our conventions.
\bigskip

\noindent
{\bf Definitions of {$\mathbf{\sigma}$ matrices and $\epsilon$ symbols}}
\bea
\sigma_{\mu} = (\mathbbm{1}_2, i \sigma^a ), ~~~ \bar{\sigma}_{\mu} = (\mathbbm{1}_2, -i \sigma^a ), ~~~ 
\varepsilon^{12} = \varepsilon_{21} = 1\, .
\eea
Note that using these conventions we are setting
\bea
\epsilon^{\alpha \beta} \ := \  i {\sigma_2}^{\alpha \beta} \ , 
\eea
whereas
\bea
\epsilon_{\alpha \beta} \ := \ - i (\sigma_2)_{\alpha \beta} \ .
\eea
It is useful to remember the basic identity
\bea
\sigma_2 \sigma^a \sigma^2 \ = \ - (\sigma^a)^T
\ . 
\eea
\bigskip

\noindent
{\bf 't Hooft  symbols}
\bea
{\eta_a}_{\mu \nu} = {\varepsilon_a}_{\mu \nu} ~~{\rm if}~~ \mu, \nu = 1,2,3, ~~~
{\eta_a}_{4 \nu} = - \delta_{a \nu} \, . 
\eea
Relation between $\eta$ and $\bar\eta$:
\bea
\bar{\eta}_{a \mu \nu} = (-1)^{\delta_{\mu 4} + \delta_{\nu 4} } \eta_{a \mu \nu}\, . 
\eea
Other useful property: 
\bea
{\eta_{a}}_{\mu \nu} {\eta^{a}}_{\kappa \lambda } = 
\delta_{\mu \kappa}\delta_{\nu \lambda} - 
\delta_{\nu \kappa}\delta_{\mu \lambda}
+ \varepsilon_{\mu \nu \kappa \lambda} \, . 
\eea
We also define 
\bea
\sigma_{\mu \nu} = i \eta_{a \mu \nu} \sigma^a,~~~
\bar{\sigma}_{\mu \nu} = i \bar{\eta}_{a \mu \nu} \bar{\sigma}^a
\ . 
\eea
With these definitions, $\eta$ and $\bar\eta$ are self-dual and anti self-dual, respectively. Likewise,  $\sigma_{\mu\nu}$  ($\bar\sigma_{\mu \nu}$) is (anti) self-dual. 

\bigskip

\noindent
{\bf Basic identities for $\sigma$ matrices}
\bea \nonumber
&&
{\rm Tr}(\sigma^a \sigma^b) = 2 \delta^{ab}, ~~ 
{\rm Tr}(\sigma^a \sigma^b \sigma^c) = 2 i\epsilon^{abc},~~ 
{\rm Tr}(\sigma^a \sigma^b \sigma^c\sigma^d) 
= 2 (\delta^{ab} \delta^{cd} - \delta^{ac} \delta^{bd} + \delta^{ad}\delta^{bc}),
 \\ \nonumber \cr
&& {\rm Tr}(\sigma^a \sigma^b \sigma^c \sigma^d \sigma^e) 
= 2i ( \epsilon^{abc} \delta^{de} + 
\epsilon^{cde}\delta^{ab}  - \epsilon^{bde}\delta^{ac} + \epsilon^{ade}\delta^{bc} ).
\eea

\bigskip

\noindent
{\bf Basic identities for Grassmann spinors}
\bea
&& \nonumber
\psi^{\alpha} = \varepsilon^{\alpha \beta } \psi_{\beta}, ~~~ 
\psi_{\alpha} = \varepsilon_{\alpha \beta } \psi^{\beta},~~~  
\psi \chi = \psi^{\alpha} \chi_{\alpha} = - \psi_{\alpha} \chi^{\alpha} = \chi^{\alpha} \psi_{\alpha}=\chi \psi
 \, , \\ \nonumber \cr
&& \bar{\psi}^{\dot{\alpha}} = \varepsilon^{\dot{\alpha} \dot{\beta} } \bar{\psi}_{\dot{\beta}}, ~~~ 
\bar{\psi}_{\dot{\alpha}} = \varepsilon_{\dot{\alpha} \dot{\beta} } \bar{\psi}^{\dot{\beta}},~~~  
\bar{\psi} \bar{\chi} = \bar{\psi}_{\dot{\alpha}} \bar{\chi}^{\dot{\alpha}} 
= - \bar{\psi}^{\dot{\alpha}} \bar{\chi}_{\dot{\alpha}} 
= \bar{\chi}_{\dot{\alpha}} \bar{\psi}^{\dot{\alpha}}= \bar{\chi} \bar{\psi} 
\, .
\eea

\bigskip

\noindent
{\bf Additional useful identities for Grassmann spinors}

\bea \nonumber
&& \bar{\sigma}^{\dot{\alpha}\alpha}_{\mu} = 
\varepsilon^{\dot{\alpha} \dot{\beta} } \varepsilon^{\alpha \beta }  {\sigma_{\mu}}_{\beta \dot{\beta}},
~~~ \varepsilon^{\alpha \beta }  \varepsilon_{\beta \gamma } = {\delta^{\alpha}}_{\gamma},
\\ \nonumber \cr
&&
\psi^{\alpha} \psi^{\beta} = - {1 \over 2} \psi^2 \varepsilon^{\alpha \beta },~~~
\psi_{\alpha} \psi_{\beta} = {1 \over 2} \psi^2 \varepsilon_{\alpha \beta },
~~~ \bar{\psi}_{\dot{\alpha} } \bar{\psi}_{\dot{\beta}} 
= - {1 \over 2} \bar{\psi}^2 \varepsilon_{\dot{\alpha} \dot{\beta} }, 
\\ \nonumber \cr
&&
\bar{\psi}^{\dot{\alpha} } \bar{\psi}^{\dot{\beta}} 
=  {1 \over 2} \bar{\psi}^2 \varepsilon^{\dot{\alpha} \dot{\beta} },  ~~~
(\eta \psi)(\eta \chi) = -{1 \over 2} (\psi \chi) \eta^2, ~~~
(\bar{\eta} \bar{\psi})(\bar{\eta} \bar{\chi}) 
= -{1 \over 2} (\bar{\psi} \bar{\chi}) \bar{\eta}^2,  \\ \nonumber \cr
&&
\psi \sigma_{\mu} \bar{\chi} = - \bar{\chi} \bar{\sigma}_{\mu} \psi, ~~~
\psi \sigma^{a} \chi = - \chi \sigma^a \psi \, , ~~~
\psi_{\alpha} \chi^{\beta}  = {1 \over 2} (- {\delta_{\alpha}}^{\beta} \psi \chi +  {(\sigma^c)_{\alpha}}^{\beta} \psi \sigma_c \chi )\, .
\eea 


\section{Instanton paraphernalia} 
\label{Appendix:instanton}
The instanton gauge connection can be efficiently expressed using the ADHM construction  \cite{Atiyah:1978ri} (for reviews see for instance \cite{Amati:1988ft,Dorey:2002ik,Bianchi:2007ft}).
It is given by the following expression: 
\beq
A \ = \ u^\dagger d u\, , 
\eeq
where, for gauge group $SU(2)$, $u$ is a $(k+1)\times 1$ matrix with quaternionic entries which are determined in the following way. 
One introduces a $(k+1)\times k$ quaternionic matrix that is linear in $x:= x_\mu \sigma_\mu$,
\beq
\Delta\, =\, a\, +\, b\, x \ .  
\eeq
Then $u$ must satisfy
\beq
\Delta^\dagger  u \, = \, 0 \, , 
\eeq
and 
\beq
\label{udu} 
u^\dagger u \, = \, \mathbbm{1}_2 \ . 
\eeq
In other words, $u$ defines an orthonormal basis of $Ker \, \Delta^\dagger$.
The condition that $F_{\mu \nu}$  is self-dual is imposed by the further requirement that 
$\Delta$ obeys 
\beq
\Delta^\dagger\Delta=f^{-1}\otimes\mathbbm{1}_2\, ,
\eeq
where  $f$ is a   $k\times k$ real matrix.   

Next we briefly discuss the symmetries of the ADHM construction. 
Gauge symmetry is realised as right-multiplication  of the matrix $u$ by a unitary quaternion $V$, $u\to u V$.  
Indeed, under this transformation 
\be
A \to \ V^\dagger A V + V^\dagger d V
\ . 
\ee
Besides gauge symmetry, we still have the freedom to transform $\Delta$ as 
\be
\label{freedom}
\Delta\to Q\Delta R \ ,
\ee 
with   $Q\in Sp(k+1)$ and $R\in GL(k,\mathbb{R})$. These symmetries can be used to simplify the expression of  $a$ and   $b$.  Specifically, $b$  can be put in ``canonical form", 
 \be
b \, =\, -
\begin{pmatrix} 0_{1\times k}\cr\mathbbm{1}_{k\times k}
\end{pmatrix} 
\ . 
\label{boh}
\ee
In this way, all instanton moduli will appear in the matrix $a$. 
For instanton number $k=1$ we choose $a$ to be of the form 
\be
a \ = \ 
\begin{pmatrix}
q \cr x_0
\end{pmatrix} \ , 
\ee
so that 
\be
\Delta = 
\begin{pmatrix}
q \cr x_0 - x
\end{pmatrix} \ . 
\ee
Physically,  $|q| := \rho $ describes the instanton size and $q/|q|$ its orientation in colour space. $x_0$ corresponds to the position of the instanton centre. 

The matrix $u$ defining the gauge connection corresponding to the choices of $a$ and $b$ given above is, up to a gauge transformation,  
\be
u \ = \ {1\over \rho \sqrt{ \rho^2 + (x-x_0)^2 }} 
\begin{pmatrix}
q (x^\dagger - x^\dagger_0) \cr\cr \rho^2
\end{pmatrix} \ . 
\ee
A short calculation leads to the explicit form of $A$, given by 
\be
A^{(0)}_\mu \ = \ - {\sigma_{\mu \nu} (x-x_0)^{\nu} \over \rho^2 + (x - x_0)^2} 
 \ . 
 \ee
 This is often called the instanton in the non-singular gauge. 
Finally we can derive the explicit form of the curvature, which is given, for any $k$, by the compact formula \cite{Osborn:1981yf} 
\be
\label{curvF}
F^{(0)}_{\mu \nu} \ = \ 2 (u^\dagger b f \sigma_{\mu \nu} b^\dagger u)
\ . 
\ee
Note that \eqref{curvF} is manifestly self-dual and gauge covariant. For $k=1$ and for our choice of $u$ we obtain
\be
F^{(0)}_{\mu \nu} \ = \ {2 \rho^2  \over \left[ \rho^2 + (x-x_0)^2 \right]^2}\, \sigma_{\mu \nu}
\ . 
 \ee

\section{Embedding space formalism} 
\label{appendix:mellin}

In order to study correlation functions in CFT, and Witten diagrams in the context of AdS/CFT correspondence, it is often convenient to consider the embedding space formalism, which has the advantage of making conformal symmetry manifest. In the following we briefly review key facts about this formalism. 

\subsection{Generalities}
In the embedding space formalism, $d$-dimensional conformal symmetry is realised as an isometry of $M_{d+2}$, 
namely $SO(d, 2)$, for more details see for instance \cite{Dirac, Mack:2009mi, Mack:2009gy, weinberg, Penedones:2010ue, Fitzpatrick:2011ia,Paulos:2011ie}. 
It is  an embedding space for AdS$_{d+1}$ with metric
\bea
ds^2 = - dX^{+} dX^{-} + \delta_{m, n} dx^{m} dx^{n}\ .
\eea
AdS$_{d+1}$ bulk coordinate $X$ and boundary coordinates $P$ can  be parameterised as
\bea
\label{hulk}
X^{A} \ = \  {1 \over \rho} (1, \rho^2 + x_0^2, x_0^{\mu}) \ , \qquad 
P^{A} \ =  \ (1, x^2, x^{\mu})\ , 
\eea
where we have set AdS radius $R = 1$. Thus we get%
\footnote{We note that  if $X:= (X^+, X^-, X^\mu)$, then our conventions are such that 
$X\cdot Y := -{1\over 2} (X^+ Y^- + X^- Y^+) + X^\mu Y_\mu$.} 
\beq
X^2 \ =\ -1 \ , \qquad P_i^2 \ = \ 0 \ .  
\eeq
Furthermore we can write the following useful relations   between usual $d$-dimensional coordinates and the embedding space coordinates: 
\beqa
P^2_{ij} & = & -2P_i \cdot P_j \ = \ x^2_{i,j} \ , 
\nonumber \\
-2P \cdot X &=& {1 \over \rho} \Big[\rho^2 + (x-x_0)^2\Big] \ . 
\eeqa
To go  from $(d + 2)$-dimensional tensors to $d$-dimensional ones, we need a pull-back
\bea 
\label{pullback}
v^{A}_{\mu}(P) \ = \ {\partial P^{A}(x^{\mu}) \over \partial x^{\mu} }\ .
\eea
Note that $v^{A}_{\mu}(P)$ is independent of $\rho$ and $x_0$, and 
\beqa
v^{A}_{\mu}(P) X_A & = & {x_{\mu}-{x_0}_{\mu} \over \rho}\ , 
\nonumber \\
v^{A}_{\mu}(P_i) {P_j}_A & = & {x_i}_{\mu} - {x_j}_{\mu} \ .
\eeqa

\subsection{Propagators in embedding space and Mellin integrals} 

In the embedding space formalism, a bulk-to-boundary propagator of a field with conformal dimension $\Delta$ can be written as
\beq
K_{\Delta}(x; z) = {1 \over (-2 P \cdot X)^{\Delta}} 
\ = \ {1 \over \Gamma(\Delta) } \int^{+\infty }_{0}\!{dt \over t}\  t^{\Delta} \, e^{2 t P \cdot X} \, , 
\eeq
where the bulk coordinate $X$ and the boundary coordinate $P$ are defined in \eqref{hulk}. 
It is then straightforward to see that 
\bea
K_{\Delta}(x; z) \ = \ {1 \over (-2 P \cdot X)^{\Delta}} =\left( { \rho \over \rho^2 + (x-x_0)^2  } \right)^{\Delta} \, . 
\eea
In some of the instanton calculations performed earlier, we have encountered derivatives acting on objects such as $\log \big( K_1(x; z)\big)$. In the embedding space, this quantity is expressed as 
\bea
\partial_{x^{\mu}} \log  ( K_1 (x; z) )
&=& {2v^{A}(P)_{\mu} X_{A} \over -2(P \cdot X)}\  =   \ 
2 v^{A}_{\mu}(P) X_A \int^{+\infty }_{0}\!{dt \over t} 
\ t \, e^{2t P \cdot X }
\\ \nonumber \cr
&=& 
 v^{A}_{\mu}(P) \partial_{P^A} \int^{+\infty }_{0}\!{dt \over t} \ t^{\varepsilon } \, e^{2t P \cdot X }
\ = \ 
 \partial_{x^{\mu}} \int^{+\infty }_{0}\!{dt \over t} \ t^{\epsilon } \, e^{2t P \cdot X } \, ,
\eea
where $v^{A}(P)_{\mu}$ is a pull-back, see \eqref{pullback}. To give meaning to this otherwise ill-defined integral, we have added a small positive exponent $\epsilon$ to the $1/t$ term in order to regulate the integration. At the end of the calculation we are free to take the limit  $\epsilon\to0$.

\subsection{Generic tree diagrams in supergravity}

Generic tree-level Witten diagrams in the  scalar sector can be computed by using the so-called Feynman rules in Mellin space
\cite{Fitzpatrick:2011ia, Paulos:2011ie, Nandan:2011wc}. 
The case of interest to us is essentially the simplest Witten diagram, namely a contact term without any bulk-to-bulk propagators,
\bea \nonumber
{\cal G}_n &=& \int_{AdS_{d+1}} K_{\Delta_1}(x_1; z) \ldots K_{\Delta_n}(x_n; z)  \ = \ 
\Big(\prod_i {1 \over \Gamma(\Delta_i) } \Big) \int^{+ \infty }_{0} \! dX {dt_i \over t_i} t^{\Delta_i}_i e^{2 (\sum t_i P_i) \cdot X}
\\ \nonumber \cr
& = & \Big(\prod_i {1 \over \Gamma(\Delta_i) } \Big)  \pi^h \Gamma\left( {\sum_i \Delta_i - d \over 2 }\right)  \int^{+ \infty }_{0} {dt_i \over t_i}\, t^{\Delta_i}_i \, 
e^{- \sum t_i t_j P_{ij}  }
\\  \cr
& = &
 \Big(\prod_i {1 \over \Gamma(\Delta_i) } \Big)  { \pi^h \over 2   } \Gamma\left( {\sum_i \Delta_i - d \over 2 }\right)    
 \int^{c + i\infty }_{c - i\infty }  [d \alpha] \ \prod_{i<j} P^{-\alpha_{i,j}}_{ij} \Gamma (\alpha_{i,j} ) \ , 
\eea
where $h=d/2$ and we have applied Symanzik's star formula reviewed below,  
in order to obtain the final result.
$c$ is a small positive number that specifies the integration contour, and the measure $[d \alpha]$ indicates 
integration over the $n(n-3)/2$ independent variables which are left after solving $2n$ constrains for the  $n(n+1)/2$ symmetric 
$\alpha_{i,j}$ variables. The constrains are
\bea
\alpha_{i,i}=0\ , \qquad \sum_{j} \alpha_{i,j} = \Delta_i 
\ . 
\eea
So we have,
\bea
[d \alpha] = {d \alpha_{1,2} \over 2 \pi i }  {d \alpha_{1,3} \over 2 \pi i } \cdots \, ,
\eea
for any  $n(n-3)/2$ independent $\alpha_{i,j}$'s.

Finally,  we quote here for completeness Symanzik's star formula \cite{Symanzik}: 
\bea
\label{tis}
\int^{\infty }_{0}\!{dt_i \over t_i} \ t^{\Delta_i}_i \, e^{ - \sum_{ij} t_i t_j P_{ij} } 
\,  = \, 
{1 \over 2  } \int^{c + i\infty }_{c - i\infty } [d \alpha] \prod_{i <j } P^{-\alpha_{i,j}}_{ij} \Gamma (\alpha_{i,j} )
\ .
\eea
This identity can be derived by replacing some (in fact $n(n-3)/2$) exponentials using their Mellin representation, 
\bea
e^{-z} \  = \ \int^{c + i \infty }_{c - i \infty }\!{ds \over 2 \pi i} \ \Gamma(s) z^{-s} \ . 
\eea
Integration over the  $t_i$'s  in \eqref{tis} leads to the final result. 


\section{A partial non-renormalisation result}  \label{appendix:partialNR}

Correlation functions of the half-BPS scalar operators $\cO^{IJ}_{\bf 20'} := \Tr (\phi^I \phi^J) - (\delta^{IJ}/6) \Tr (\phi^L \phi^L)$ (with $I, J=1, \ldots 6$) can be efficiently repackaged in terms of the correlation functions of the operator 
\bea 
\mathcal{O}(x, Y) \, :=  \, Y^{I} Y^{J} {\rm Tr}(\phi^{I} \phi^J ) \ , 
\eea 
where $Y^{I}$
are  auxiliary $SO(6)$ harmonic null variables, with  $Y^I Y_I=0$ \cite{paul}. 
 We are interested in the four-point correlation function  in the one-instanton background, which  is given by
\bea
{\cal G}^{1-\rm{inst}}(x_i, Y_i) &=& \langle \mathcal{O}(x_1,Y_1)\mathcal{O}(x_2,Y_2)\mathcal{O}(x_3,Y_3)\mathcal{O}(x_4,Y_4) \rangle_{K=1}  \nonumber \\
&=& R(1,2,3,4) {3^4 \pi^2 \Gamma( 6 ) \over 2^5 \Gamma(4)^4 x^2_{1,3} x^2_{2,4}}  \\
&\times & \int^{+ i \infty }_{ - i \infty } {d \alpha_{3,4} \over 2 \pi i}  { d \alpha_{1,4}  \over 2 \pi i}
u^{1-\alpha_{3,4}} v^{1-\alpha_{1,4}}  
\Gamma^2(\alpha_{3,4})\Gamma^2(\alpha_{1,4}) \Gamma^2(4 - \alpha_{1,4} - \alpha_{3,4}) \,  , \nonumber
\eea
where the function $R(1,2,3,4)$ is given by \cite{ paul,Dolan:2004mu} 
\bea
R(1,2,3,4)&=&{y^2_{12} y^2_{23} y^2_{34} y^2_{41} \over x^2_{1,2} x^2_{2,3} x^2_{3,4} x^2_{4,1}}
( x^2_{1,3} x^2_{2,4} - x^2_{1,2}x^2_{3,4} - x^2_{1,4}x^2_{2,3} ) \cr
&+& {y^2_{12} y^2_{13} y^2_{24} y^2_{34} \over x^2_{1,2} x^2_{1,3} x^2_{2,4} x^2_{3,4}}
( x^2_{1,4} x^2_{2,3} - x^2_{1,2}x^2_{3,4} - x^2_{1,3}x^2_{2,4} ) \cr
&+& 
{y^2_{13} y^2_{14} y^2_{23} y^2_{24} \over x^2_{1,3} x^2_{1,4} x^2_{2,3} x^2_{2,4}}
( x^2_{1,2} x^2_{3,4} - x^2_{1,4}x^2_{2,3} - x^2_{1,3}x^2_{2,4} ) \cr
&+& {y^4_{12} y^4_{34} \over x^2_{1,2} x^2_{3,4}} + {y^4_{13} y^4_{24} \over x^2_{1,3} x^2_{2,4}}
+ {y^4_{14} y^4_{23} \over x^2_{1,4} x^2_{2,3}} \, , 
\eea
and  $Y_i \cdot Y_j := y^2_{ij}$. 

Remarkably, $R(1,2,3,4)$  has $S_4$ permutation symmetry, which implies  that the Mellin integral with the prefactor $1/(x^2_{1,3} x^2_{2,4})$ enjoys the same $S_4$ symmetry, which is straightforward to check. 
The fact that  the four-point correlators ${\cal G}^{1-\rm{inst}}(x_i, Y_i)$ can be written in this compact form is guaranteed by a partial non-renormalisation theorem of the stress-tensor multiplet correlation functions \cite{Eden:2000bk}.



\begin{thebibliography}{99}

\bibitem{Alday:2007hr} 
  L.~F.~Alday and J.~M.~Maldacena,
{\it Gluon scattering amplitudes at strong coupling,}
  JHEP {\bf 0706}, 064 (2007)
{\tt  [arXiv:0705.0303 [hep-th]].}


\bibitem{Drummond:2007aua} 
  J.~M.~Drummond, G.~P.~Korchemsky and E.~Sokatchev,
  {\it Conformal properties of four-gluon planar amplitudes and Wilson loops,}
  Nucl.\ Phys.\ B {\bf 795}, 385 (2008)
  {\tt [arXiv:0707.0243 [hep-th]].}

\bibitem{Brandhuber:2007yx} 
  A.~Brandhuber, P.~Heslop and G.~Travaglini,
  {\it MHV amplitudes in N=4 super Yang-Mills and Wilson loops,}
  Nucl.\ Phys.\ B {\bf 794}, 231 (2008)
  {\tt [arXiv:0707.1153 [hep-th]].}
  
  %
\bibitem{Alday:2010zy}
  L.~F.~Alday, B.~Eden, G.~P.~Korchemsky, J.~Maldacena and E.~Sokatchev,
  {\it From correlation functions to Wilson loops,}
  JHEP {\bf 1109} (2011) 123
 {\tt  [arXiv:1007.3243 [hep-th]].}
  
\bibitem{Eden:2010zz} 
  B.~Eden, G.~P.~Korchemsky and E.~Sokatchev,
  {\it From correlation functions to scattering amplitudes,}
  JHEP {\bf 1112}, 002 (2011)
  {\tt [arXiv:1007.3246 [hep-th]].}
  
\bibitem{Mason:2010yk} 
  L.~J.~Mason and D.~Skinner,
  {\it The Complete Planar S-matrix of N=4 SYM as a Wilson Loop in Twistor Space,}
  JHEP {\bf 1012}, 018 (2010)
{\tt   [arXiv:1009.2225 [hep-th]].}

\bibitem{CaronHuot:2010ek} 
  S.~Caron-Huot,
  {\it Notes on the scattering amplitude / Wilson loop duality,}
  JHEP {\bf 1107}, 058 (2011)
{\tt   [arXiv:1010.1167 [hep-th]].}
  
\bibitem{Eden:2011yp} 
  B.~Eden, P.~Heslop, G.~P.~Korchemsky and E.~Sokatchev,
  {\it The super-correlator/super-amplitude duality: Part I,}
  Nucl.\ Phys.\ B {\bf 869}, 329 (2013)
{\tt   [arXiv:1103.3714 [hep-th]].}
  
\bibitem{Eden:2011ku} 
  B.~Eden, P.~Heslop, G.~P.~Korchemsky and E.~Sokatchev,
  {\it The super-correlator/super-amplitude duality: Part II,}
  Nucl.\ Phys.\ B {\bf 869}, 378 (2013)
  {\tt [arXiv:1103.4353 [hep-th]].}
  
  
\bibitem{Bianchi:1998nk} 
  M.~Bianchi, M.~B.~Green, S.~Kovacs, G.~Rossi,
  {\it Instantons in supersymmetric Yang-Mills and D instantons in IIB superstring theory,}
  JHEP {\bf 9808}, 013 (1998)
  {\tt [hep-th/9807033].}
  
\bibitem{Alday:2013cwa} 
  L.~F.~Alday and A.~Bissi,
  JHEP {\bf 1310}, 202 (2013)
  [arXiv:1305.4604 [hep-th]].


  %
\bibitem{Dorey:1999pd}
  N.~Dorey, T.~J.~Hollowood, V.~V.~Khoze, M.~P.~Mattis and S.~Vandoren,
  {\it Multi-instanton calculus and the AdS / CFT correspondence in N=4 superconformal field theory,}
  Nucl.\ Phys.\ B {\bf 552} (1999) 88
 {\tt  [hep-th/9901128].}


\bibitem{Green:1997tv}
  M.~B.~Green and M.~Gutperle,
  {\it Effects of D instantons,}
  Nucl.\ Phys.\ B {\bf 498} (1997) 195
 {\tt  [hep-th/9701093].}

\bibitem{Banks:1998nr}
  T.~Banks and M.~B.~Green,
  {\it Nonperturbative effects in AdS in five-dimensions x S**5 string theory and d = 4 SUSY Yang-Mills,}
  JHEP {\bf 9805} (1998) 002
 {\tt  [hep-th/9804170].}


%
\bibitem{Basso:2007wd}
  B.~Basso, G.~P.~Korchemsky and J.~Kotanski,
  {\it Cusp anomalous dimension in maximally supersymmetric Yang-Mills theory at strong coupling,}
  Phys.\ Rev.\ Lett.\  {\bf 100} (2008) 091601
 {\tt  [arXiv:0708.3933 [hep-th]].}
  
\bibitem{Polyakov:1980ca}
A.~M.~Polyakov,
{\it Gauge Fields As Rings Of Glue},
Nucl.\ Phys.\  B {\bf 164} (1980) 171.

\bibitem{Brandt:1981kf}
  R.~A.~Brandt, F.~Neri and M.~a.~Sato,
  {\it Renormalization Of Loop Functions For All Loops,}
  Phys.\ Rev.\  D {\bf 24} (1981) 879.

\bibitem{Korchemsky:1985xj}
  G.~P.~Korchemsky and A.~V.~Radyushkin,
  {\it Loop Space Formalism And Renormalization Group For The Infrared Asymptotics Of QCD,}
  Phys.\ Lett.\  B {\bf 171} (1986) 459.

  
  
    \bibitem{LSZ}
  M.~Bianchi, A.~Brandhuber, G.~Travaglini and C.~Wen,
  in preparation. 

\bibitem{SW}
N.~Seiberg and E.~Witten,
{\it Electric - magnetic duality, monopole condensation, and confinement in N=2 supersymmetric Yang-Mills theory,}
  Nucl.\ Phys.\ B {\bf 426} (1994) 19
   [Erratum-ibid.\ B {\bf 430} (1994) 485]
{\tt  [hep-th/9407087].}

\bibitem{PF}
D.~Finnell and P.~Pouliot,
{\it Instanton calculations versus exact results in four-dimensional SUSY gauge theories,}
  Nucl.\ Phys.\ B {\bf 453} (1995) 225
{\tt  [hep-th/9503115].}

\bibitem{Green:2002vf}
  M.~B.~Green and S.~Kovacs,
  {\it Instanton induced Yang-Mills correlation functions at large N and their AdS(5) x S**5 duals,}
  JHEP {\bf 0304} (2003) 058
{\tt   [hep-th/0212332].}


\bibitem{Kovacs:2003rt}
  S.~Kovacs,
  {\it On instanton contributions to anomalous dimensions in N=4 supersymmetric Yang-Mills theory,}
  Nucl.\ Phys.\ B {\bf 684} (2004) 3
  {\tt [hep-th/0310193].}
  
  
\bibitem{Bianchi:1999ge} 
  M.~Bianchi, S.~Kovacs, G.~Rossi and Y.~S.~Stanev,
 {\it On the logarithmic behavior in N=4 SYM theory,}
  JHEP {\bf 9908}, 020 (1999)
  {\tt [hep-th/9906188].}
  
  
  
  
  
  
  
  
\bibitem{Corrigan:1978xi}
  E.~Corrigan, P.~Goddard and S.~Templeton,
  {\it Instanton Green's Functions and Tensor Products,}
  Nucl.\ Phys.\ B {\bf 151} (1979) 93.
  

\bibitem{hooft}
  G.~'t Hooft,
  {\it Computation of the Quantum Effects Due to a Four-Dimensional Pseudoparticle,}
  Phys.\ Rev.\ D {\bf 14} (1976) 3432
   [Erratum-ibid.\ D {\bf 18} (1978) 2199].


%
\bibitem{Penedones:2010ue} 
  J.~Penedones,
  {\it Writing CFT correlation functions as AdS scattering amplitudes,}
  JHEP {\bf 1103}, 025 (2011)
{\tt   [arXiv:1011.1485 [hep-th]].}

%
\bibitem{Fitzpatrick:2011ia} 
  A.~L.~Fitzpatrick, J.~Kaplan, J.~Penedones, S.~Raju and B.~C.~van Rees,
  {\it A Natural Language for AdS/CFT Correlators,}
  JHEP {\bf 1111}, 095 (2011)
 {\tt  [arXiv:1107.1499 [hep-th]].}
  
\bibitem{Paulos:2011ie} 
  M.~F.~Paulos,
  {\it Towards Feynman rules for Mellin amplitudes,}
  JHEP {\bf 1110}, 074 (2011)
{\tt   [arXiv:1107.1504 [hep-th]].}
  
  
  
  \bibitem{Nandan:2011wc} 
  D.~Nandan, A.~Volovich and C.~Wen,
  { \it On Feynman Rules for Mellin Amplitudes in AdS/CFT, }
  JHEP {\bf 1205}, 129 (2012)
  [arXiv:1112.0305 [hep-th]].


\bibitem{SVP}
M.~F.~Paulos, M.~Spradlin and A.~Volovich,
 {\it Mellin Amplitudes for Dual Conformal Integrals,}
  JHEP {\bf 1208} (2012) 072
{\tt  [arXiv:1203.6362 [hep-th]].}

 %
  \bibitem{Bianchi:2007ft}
  M.~Bianchi, S.~Kovacs, G.~Rossi and ,
  {\it Instantons and Supersymmetry,}
  Lect.\ Notes Phys.\  {\bf 737} (2008) 303
{\tt   [hep-th/0703142 [hep-th]].}
  
  
  \bibitem{Atiyah:1978ri}
  M.~F.~Atiyah, N.~J.~Hitchin, V.~G.~Drinfeld, Y.~I.~Manin
  {\it Construction of Instantons,}
  Phys.\ Lett.\ A {\bf 65} (1978) 185.

  
  \bibitem{Bianchi:2001cm}
  M.~Bianchi, S.~Kovacs, G.~Rossi and Y.~S.~Stanev,
  {\it Properties of the Konishi multiplet in N=4 SYM theory,}
    JHEP {\bf 0105} (2001) 042
  {\tt [hep-th/0104016].}
    

\bibitem{Amati:1988ft}
  D.~Amati, K.~Konishi, Y.~Meurice, G.~C.~Rossi, G.~Veneziano and ,
 {\it Nonperturbative Aspects in Supersymmetric Gauge Theories,}
  Phys.\ Rept.\  {\bf 162} (1988) 169.
  
\bibitem{Dorey:2002ik}
  N.~Dorey, T.~J.~Hollowood, V.~V.~Khoze and M.~P.~Mattis, 
  {\it The Calculus of many instantons,}
  Phys.\ Rept.\  {\bf 371} (2002) 231
  {\tt [hep-th/0206063].}
  
  %
\bibitem{Osborn:1981yf}
  H.~Osborn,
  {\it Semiclassical Functional Integrals For Selfdual Gauge Fields,}
  Annals Phys.\  {\bf 135} (1981) 373.
  
  \bibitem{Dirac}
   P. A. M. Dirac, 
   {\it Wave equations in conformal space,}
   Ann. Math.\ 37, 429 (1936)
  


\bibitem{Mack:2009mi}
  G.~Mack,
  {\it D-independent representation of Conformal Field Theories in D dimensions via transformation to auxiliary Dual Resonance Models. Scalar amplitudes,}
{\tt  arXiv:0907.2407 [hep-th].}

\bibitem{Mack:2009gy}
  G.~Mack,
  {\it D-dimensional Conformal Field Theories with anomalous dimensions as Dual Resonance Models,}
  Bulg.\ J.\ Phys.\  {\bf 36} (2009) 214
{\tt   [arXiv:0909.1024 [hep-th]].}

  \bibitem{weinberg}
  S.~Weinberg,
{\it Six-dimensional Methods for Four-dimensional Conformal Field Theories,}
  Phys.\ Rev.\ D {\bf 82} (2010) 045031
 {\tt  [arXiv:1006.3480 [hep-th]].}
  
  \bibitem{Symanzik}
K.~Symanzik,
 {\it On Calculations in conformal invariant field theories,}
Lett.Nuovo Cim. 3 (1972) 734-738


 \bibitem{paul}
  B.~Eden, P.~Heslop, G.~P.~Korchemsky and E.~Sokatchev,
  {\it Hidden symmetry of four-point correlation functions and amplitudes in N=4 SYM,}
  Nucl.\ Phys.\ B {\bf 862} (2012) 193
  {\tt [arXiv:1108.3557 [hep-th]].}
  
\bibitem{Dolan:2004mu}
F.~A.~Dolan, L.~Gallot and E.~Sokatchev,
{\it On four-point functions of 1/2-BPS operators in general dimensions,}
JHEP {\bf 0409}, 056 (2004)
{\tt [hep-th/0405180].}
 

  
 
\bibitem{Eden:2000bk} 
  B.~Eden, A.~C.~Petkou, C.~Schubert and E.~Sokatchev,
  {\it Partial nonrenormalization of the stress tensor four point function in N=4 SYM and AdS / CFT,}
  Nucl.\ Phys.\ B {\bf 607}, 191 (2001)
 {\tt  [hep-th/0009106].}

 


\end{thebibliography}
\end{document}